\documentclass[12pt,journal,onecolumn,draftcls]{IEEEtran}
\usepackage{epsfig}
\usepackage{times}
\usepackage{float}
\usepackage{afterpage}
\usepackage{amsmath}
\usepackage{amstext}
\usepackage{amssymb,bm}
\usepackage{latexsym}
\usepackage{color}
\usepackage{graphicx}
\usepackage{amsmath}
\usepackage{amsthm}
\usepackage{graphicx}
\usepackage[center]{caption}
\usepackage[normalem]{ulem}
\usepackage{pstricks}
\usepackage{subfigure}
\usepackage{booktabs}
\usepackage{multicol}
\usepackage{lipsum}
\usepackage{dblfloatfix}
\usepackage{cancel}

\allowdisplaybreaks

\newcommand{\mcal}{\mathcal}

\newtheorem{thm}{Theorem}
\newtheorem{cor}[thm]{Corollary}
\newtheorem{lem}[thm]{Lemma}

\newtheorem{rem}{Remark}
\newtheorem{defin}{Definition}
\newtheorem{prope}{Property}

\begin{document}
\title{\vspace{0.25in} Network Simplification in Half-Duplex: \\
Building on Submodularity}
\author{Martina~Cardone$^\star$, Yahya~H.~Ezzeldin$^\star$, Christina~Fragouli and Daniela~Tuninetti\\
$^\star$ Co-First Authors
\thanks{
M.~Cardone, Y.~H.~Ezzeldin and C.~Fragouli are with the Electrical Engineering Department at the University of California, Los Angeles, CA 90095 USA (e-mail: \{martina.cardone, yahya.ezzeldin, christina.fragouli\}@ucla.edu). 
The research carried out at UCLA was partially funded by NSF under award number 1514531 and 1314937.

D. Tuninetti is with the Electrical and Computer Engineering Department of the University of Illinois at Chicago, Chicago, IL 60607 USA (e-mail: danielat@uic.edu). 
The work of D.~Tuninetti was partially funded by NSF under award number 1527059.

The results in this paper were presented in part at the 2016 IEEE International Symposium on Information Theory.
}
}

\maketitle
\begin{abstract}
This paper explores the {\it network simplification} problem in the context of Gaussian Half-Duplex (HD) diamond networks.
Specifically, given an $N$-relay diamond network, this problem seeks to derive fundamental guarantees on the capacity of the best $k$-relay subnetwork, as a function of the full network capacity.
The main focus of this work is on the case when $k=N-1$ relays are selected out of the $N$ possible ones.
First, a simple algorithm, which removes the relay with the minimum capacity (i.e., the worst relay), is analyzed and it is shown that the remaining $(N-1)$-relay subnetwork has an approximate (i.e., optimal up to a constant gap) HD capacity that is at least half of the approximate HD capacity of the full network.
This fraction guarantee is shown to be tight if only the single relay capacities are known, i.e., there exists a class of Gaussian HD diamond networks with $N$ relays where, by removing the worst relay, the subnetwork of the remaining $k=N-1$ relays has an approximate capacity equal to half of the approximate capacity of the full network.
Next, this work proves a fundamental guarantee, which improves over the previous fraction: there always exists a subnetwork of $k=N-1$ relays that achieves at least a fraction $\frac{N-1}{N}$ of the approximate capacity of the full network.
This fraction is proved to be tight and it is shown that any optimal schedule of the full network can be used by at least one of the $N$ subnetworks of $N-1$ relays to achieve a worst-case performance guarantee of $\frac{N-1}{N}$.
Additionally, these results are extended to derive lower bounds on the fraction guarantee for general $k \in [1:N]$.
The key steps in the proofs lie in the derivation of properties of submodular functions, which provide a combinatorial handle on the network simplification problem in Gaussian HD diamond networks.
Finally, this work provides comparisons between the simplification problem for HD and Full-Duplex (FD) networks that highlight their different natures.
For instance, it is shown that in HD, different from the FD counterpart, when $k \in \{1,2\}$ the fraction guarantee decreases as $N$ increases.
\end{abstract}

\section{Introduction}
Consider a relay network where a (potentially large) number of relays assist the over-the-air communication from a source to a destination.
The wireless network simplification problem seeks to answer the following question: can a significant fraction of the capacity of the full network be achieved by operating only a subset of the available relays?

Wireless network simplification was pioneered by the authors in~\cite{NazarogluIT2014} in the context of Gaussian Full-Duplex (FD) diamond networks\footnote{An $N$-relay diamond network is a two-hop relay network where the source communicates with the destination through $N$ non-interfering relays.}.
The importance of this problem stems from the several benefits it offers.
For example, operating all the available relays might be computationally expensive as the relays must coordinate for transmission and might incur a significant cost in terms of consumed power.
Network simplification represents a potential solution to these limiting factors as it promises {\it energy savings} -- since only the power of the active relays is used to transmit information -- and a complexity reduction in the {\it synchronization} problem -- since only the selected relays have to be synchronized for transmission -- while ensuring that a significant fraction of the capacity of the full network is achieved.

In this paper, we investigate the network simplification problem for Gaussian Half-Duplex (HD) diamond networks with $N$ relays.
Our study is motivated by the fact that currently employed relays operate in HD, unless sufficient isolation between the antennas can be guaranteed or different bands are used for transmission and reception.
Additionally, as recently announced in 3GPP Rel-13, HD is also expected to be employed in next generation Internet of Things networks to enable low-cost communication modules for short-distance and infrequent data transmissions.

Studying the network simplification problem is more challenging when networks operate in HD compared to FD. This is due to the intrinsic combinatorial nature of capacity characterization in HD relay networks, as elaborated in the following summary of relevant related work.


\subsection {Related Work}
The capacity characterization of the Gaussian HD relay network is a long-standing open problem. 
The tightest upper bound on the capacity is the well-known cut-set upper bound~\cite{coveElGamal}. 
A number of schemes have been proposed~\cite{LimIT2011},~\cite{AvestimehrIT2011},~\cite{OzgurIT2013},~\cite{LimISIT2014} that achieve the cut-set upper bound to within a constant gap (independently of the channel parameters). 
To the best of our knowledge, the tightest refinement of the achievable gap is $1.96 (N+2)$ bits/sec derived in~\cite{CardoneIT2014}\footnote{The constant gap in~\cite{CardoneIT2014} was derived by using the approach first proposed in~\cite{KramerAllerton2004}. The work in~\cite{KramerAllerton2004} showed that HD relay networks can be studied within the framework of their FD counterparts, by expressing the channel inputs and outputs as functions of the states of the relays. 
In particular, it was observed that information can be conveyed by randomly switching the relay between transmit and receive modes. However, this only improves the capacity by a constant, at most $1$ bit per relay.}, where $N$ is the number of relays in the network.
Given these results, the cut-set bound evaluated with independent inputs, is said to approximate the capacity (i.e., up to a gap that only depends on $N$).
In the rest of the paper, we refer to this bound as the {\it approximate capacity}.
We also point out that, although for some specific network topologies in FD -- such as Gaussian FD diamond networks~\cite{SenguptaITW2012,ChernITW2012} -- the constant gap has been shown to grow sub-linearly with $N$, for general Gaussian relay networks a linear in $N$ gap to the cut-set bound is fundamental~\cite{OzgurISIT2015,OzgurAllerton2015}.


In general, the capacity characterization (or the evaluation of the approximate capacity) of HD relay networks is more challenging than the FD counterpart since, in addition to the optimization over the $2^N$ cuts, it also requires an optimization over the $2^N$ listen/transmit configuration states.
We refer to the states that suffice to characterize the approximate capacity by  {\em active} states.
Recently, in~\cite{CardoneITW2015} the authors proved a surprising result, which was first conjectured in~\cite{BrahmaISIT2012}: at most $N+1$ states (out of the $2^N$ possible ones) are active in the simplest optimal schedule (one with the least number of active states) for a class of HD relay networks, which includes the practically relevant Gaussian noise network.
This result generalizes those in~\cite{BagheriIT2014},~\cite{BrahmaISIT2014} and~\cite{BrahmaIT2016}, valid only for Gaussian HD relay networks with a diamond topology and limited network sizes. 
The result in~\cite{CardoneITW2015} is promising as it can lead to a significant operational complexity reduction (from operating the network with an exponential number of states in $N$ to linear in $N$). 
Furthermore, this result might be leveraged to efficiently evaluate the approximate capacity, as we recently showed in~\cite{EzzeldinISIT2017} in the context of Gaussian HD line networks. 
However, even though we understand that such a schedule exists (with at most $N+1$ active states), to the best of our knowledge, it is not yet known if we can find these states efficiently for general relay networks.
A similar thread of research~\cite{SenguptaITW2014} has focused on deriving capacity guarantees when each relay operates with its optimal schedule (computed as if the other relays were not there) and is allowed to switch multiple times between listen and transmit modes of operation.
For capacity evaluation, the authors in~\cite{EtkinParvareshShomoronyAvestimehr} proposed an approach that, for certain network topologies -- such as the line network and a specific class of layered networks -- outputs the approximate capacity in polynomial time. 
This result is quite promising, but it relies on the simplified topology of certain class of relay networks.

Different from the aforementioned thread of research, where the main objective is to provide a low-complexity characterization of the network capacity when all the $N$ relays are active, in this work, we seek to understand what fraction can be guaranteed when only a subset of $k \in [1:N]$ relays is operated.
This problem was first explored by the authors in~\cite{NazarogluIT2014} in the context of Gaussian FD diamond networks.
Specifically, the authors in~\cite{NazarogluIT2014} showed that, in any $N$-relay Gaussian FD diamond network, there always exists a subnetwork of $k$ relays that achieves at least a fraction $\frac{k}{k+1}$ of the approximate capacity of the full network.
This result, which is independent of $N$, is quite promising as it implies that a significant fraction of the approximate capacity can be achieved by operating only $k$ relays, out of the $N$ possible ones.
This fraction guarantee was proved to be tight, i.e., there exist $N$-relay Gaussian FD diamond networks for which the best $k$-relay subnetwork (i.e., the one with the largest approximate capacity) achieves this fraction of the full network approximate FD capacity.
A polynomial-time algorithm to discover these high-capacity $k$-relay subnetworks was also proposed in~\cite{NazarogluIT2014}.
Recently, in~\cite{EzzeldinISIT2016} the authors considered a more general network, namely the Gaussian FD layered network and proved a worst-case fraction guarantee for selecting 
the best path in the network. 
From the result in~\cite{NazarogluIT2014}, it directly follows that in Gaussian HD diamond networks, by selecting $k$ relays, one can always achieve at least a fraction $\frac{k}{2(k+1)}$ of the approximate HD capacity of the whole network. 
This is accomplished by operating the $k$ relays (selected as in FD) in only $2$ states (out of the $2^k$ possible ones) of equal duration: the first where all the $k$ relays listen and the second where all the $k$ relays transmit.
Although providing a performance guarantee, this result might be too conservative.
This is indeed confirmed by the result in~\cite{BrahmaISIT2014Relay} where it was proved that, in any Gaussian HD diamond network, there always exists a subnetwork of $k=2$ relays that, when operated in complementary fashion (i.e., when one relay transmits, the other listens and vice versa), achieves at least half of the approximate capacity of the full network.
In this paper, we do not restrict the selected $k$ relays to operate only in certain states as in~\cite{BrahmaISIT2014Relay}, which leads to better performance guarantees in terms of achievable fraction of the approximate capacity.

\subsection{Contributions}
In this paper we seek to understand how much of the approximate HD capacity one can achieve by smartly selecting a subset of $k$ relays out of the $N$ possible ones in a Gaussian HD diamond network.
In particular, our goal is to provide a worst-case performance guarantee (in terms of achievable fraction) that holds universally (i.e., independently of the values of the channel parameters).
Our main contributions can be summarized as follows:
\begin{enumerate}
\item We first derive properties of Gaussian diamond networks and submodular functions, which provide a combinatorial handle on the network simplification problem in Gaussian HD diamond networks.
    For instance, we prove a result {that} we refer to as the {\it partition lemma}, which states that if we partition the network into multiple subnetworks such that each relay belongs to only one of such subnetworks, then the approximate capacity of the full network is upper bounded by the sum of the approximate capacities of the subnetworks.
Beyond their utilization in the proofs of our main results, these properties might be of independent interest.
\item We analyze a straightforward algorithm to select a subnetwork of $k=N-1$ relays, which operates all the relays except the {\it worst} one. 
We say that, among the $N$ relays, the $i$-th relay is the {\it worst} if it has the smallest single approximate capacity, i.e., if the maximum HD flow that can be routed through it is less than or equal to the other $N-1$ flows through each of the remaining $N-1$ relays.
We prove that the algorithm outputs, in linear time, a subnetwork whose approximate HD capacity is at least half of the approximate HD capacity of the whole network.
We also show that this fraction guarantee is tight if we know only the single relay capacities, i.e., there exists a class of Gaussian HD diamond networks with $N$ relays where, by removing the worst relay, the remaining $(N-1)$-relay subnetwork has an approximate capacity that is half of the approximate capacity of the full network.
This guarantee might be too conservative and indeed a smarter choice leads to a better performance, as described in the next point.
However, an appealing feature of this algorithm is that it only requires the knowledge of the $N$ single capacities.
%
\item We prove that, in any $N$-relay Gaussian HD diamond network, there always exists a subnetwork of $k=N-1$ relays that achieves at least a fraction $\frac{N-1}{N}$ of the approximate capacity of the full network.
We also show that this fraction of $\frac{N-1}{N}$ is tight.
This result significantly improves over the fraction of half guaranteed by the algorithm described in the previous point. 
Moreover, this guarantee is fundamental, i.e., it is the largest fraction that can be ensured when $N-1$ relays are selected.
In addition, we show a surprising result:
any optimal schedule of the full network can be used by at least one of the $N$ subnetworks of $k=N-1$ relays to achieve the worst performance guarantee.
This leads to a complexity reduction in the scheduling problem; in fact, it implies that, in order to select an $(N-1)$-relay subnetwork that achieves a fraction $\frac{N-1}{N}$ of the approximate capacity of the full network, there is no need to compute the optimal schedule for each of the $N$ subnetworks.
It suffices to compute an optimal schedule of the full network. 

%
\item We generalize the results described in the previous two points to generic values of $k \in [1:N]$.
In particular, we show that:
(i) the straightforward algorithm that removes the $N-k$ worst relays and runs in $O(N\log(N))$, ensures that the selected $k$-relay subnetwork has an approximate capacity that is at least $2^{-(N-k)}$ of the approximate capacity of the original network with $N$ relays;
(ii) a fraction $\frac{k}{N}$ of the approximate capacity of the full network can always be achieved by selecting $k$ relays and operating them with an optimal schedule of the full network.
However, this last worst-case fraction guarantee does not appear to be tight.
This result suggests that, when $k<N-1$, forcing the $k$-relay subnetworks to operate with the optimal schedule of the full network is suboptimal.

%
\item We find significant differences between the wireless simplification problem for HD and FD networks. 
For instance:
(i) in HD, when $k \in \{1,2\}$ relays are selected, the fraction of the achieved approximate capacity depends on $N$ and decreases as $N$ increases;
(ii) the worst-case networks in HD and FD are not necessarily the same; 
(iii) the best $k$-relay subnetworks in HD and FD might be different.
These results show that FD and HD relay networks have a different nature.
This might be due to the fact that in HD the schedule plays a crucial role and hence removing some of the relays can change the schedule at which the selected subnetwork should be optimally operated.

\end{enumerate}

\subsection{Paper Organization}
Section~\ref{sec:system_model} describes the $N$-relay Gaussian HD diamond network and summarizes known capacity results.
Section~\ref{sec:Properties} derives properties of submodular functions and diamond networks. 
Section~\ref{sec:MotivatingExample} studies the performance (in terms of achievable fraction) of a simple algorithm that selects $k \in [1:N]$ relays out of the $N$ possible ones, by removing the worst $N-k$ relays.
In particular, Section~\ref{sec:MotivatingExample} first considers the case $k=N-1$ and then generalizes the result to any $k \in [1:N]$.
Section~\ref{sec:FundGar} provides a fundamental guarantee (in terms of achievable fraction) when $N-1$ relays are selected out of the $N$ possible ones.
Section~\ref{sec:FundGar} also generalizes the lower bound on the fraction guarantee for $k=N-1$ to general $k\in [1:N]$.
Finally, Section~\ref{sec:Concl} discusses some implications of the presented results, highlights differences between the selection performances in HD and FD networks and concludes the paper.
Some of the proofs can be found in the Appendix.

\subsection{Notation}
In the rest of the paper, we use the following notation convention.
We denote with $[a:b]$ the set of integers from $a$ to $b \geq a$. 
$Y^j$ is a vector of length $j$ with components $\left (Y_1,\ldots,Y_j\right )$, 
$|z|$ is the component-wise absolute value of the vector $z$ and $z^T$ is the transpose of the vector $z$.
For two sets $\mathcal{A}_1,\mathcal{A}_2$, $\mathcal{A}_1 \subseteq \mathcal{A}_2$ indicates that $\mathcal{A}_1$ is a subset of $\mathcal{A}_2$, $\mathcal{A}_1 \cup \mathcal{A}_2$ represents the union of $\mathcal{A}_1$ and $\mathcal{A}_2$, $\mathcal{A}_1 \cap \mathcal{A}_2$ represents the intersection of $\mathcal{A}_1$ and $\mathcal{A}_2$ and $\mathcal{A}_1 \backslash \mathcal{A}_2$ is the set of elements that belong to $\mathcal{A}_1$ but not to $\mathcal{A}_2$.
With $|\mathcal{A}|$ we indicate the cardinality of $\mathcal{A}$, $\emptyset$ is the empty set and $\mathbb{E}[\cdot]$ indicates the expected value.
For all $x \in \mathbb{R}$, the ceiling and floor functions are denoted by $\lceil x \rceil$ and $ \lfloor x \rfloor$, respectively. 
The $\ell_1$-norm of a vector $\lambda$ is represented by $\|\lambda\|_1$.
Table~\ref{table:Notation} summarizes and defines quantities that are frequently used throughout the paper.

\begin{table}
\caption{Quantities of interest used throughout the paper.}
\label{table:Notation}
\begin{center}
\begin{tabular}{ |c||c| }
 \hline
 {\bf{Quantity}} & {\bf{Definition}}   \\
 \hline
\hline
$\mcal{N}_{\mathcal{K}}$   & Network which contains only the relays in $\mathcal{K} \subseteq [1:N]$  \\ 
${\bar{\mathcal{N}}}_i$ & $\mathcal{N}_{[1:N] \backslash \{i\}}$ \\
${\mathsf{C}}_{\mcal{N}_{\mathcal{K}}}$ & Approximate HD capacity of $\mcal{N}_{\mathcal{K}}$ \\
${\mathsf{R}}^{\lambda}_{\mathcal{N}_{\mathcal{K}}}$ & Approximate HD achievable rate of $\mcal{N}_{\mathcal{K}}$ when operated with the schedule $\lambda$\\
$\mathsf{C}_{\mcal{N}_{\mathcal{K}}}^{\rm{FD}}$ &  Approximate FD capacity of $\mcal{N}_{\mathcal{K}}$ \\
 \hline
\end{tabular}
\end{center}
\vspace{-7mm}
\end{table}

\section{System Model and Known Results} 
\label{sec:system_model}

\begin{figure}
\centering
\includegraphics[width=0.5\textwidth]{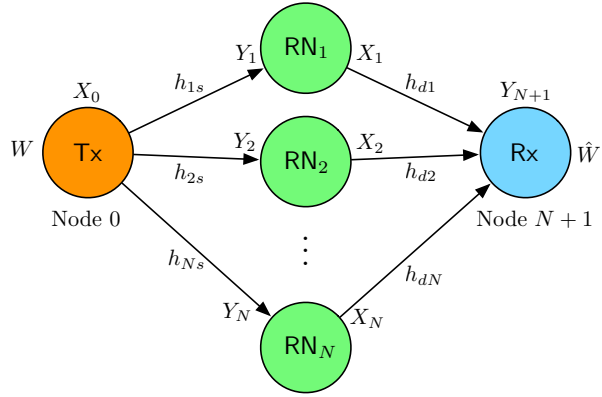}
\caption{Gaussian diamond network with $N$ relays.}
\label{fig:diamond2relays}
\end{figure}

We consider the Gaussian HD diamond network $\mcal{N}_{[1:N]}$ in Fig.~\ref{fig:diamond2relays} where a source node (node~$0$) wishes to communicate with a destination (node~$N+1$) through $N$ non-interfering relays operating in HD.
Specifically, the source has a message $W$ uniformly distributed on $\left [1:2^{KR} \right]$ for the destination, where $K \in \mathbb{N}$ denotes the codeword length and $R \in \mathbb{R}_+$ is the transmission rate in bits per channel use.
At time $t \in [1:K]$, the source maps the message $W$ into a channel input $X_{0,t} \left( W\right)$ and the $i$-th relay, with $i\in[1:N]$,  if in transmission mode of operation, maps its past channel observations into a channel input symbol $X_{i,t}\left(Y^{t-1}_i \right)$. 
At time $K$, the destination outputs an estimate $\hat{W}$ of the message based on all its channel observations $Y_{N+1}^K$.
A rate $R$ is said to be $\epsilon$-achievable if there exists a sequence of codes indexed by the block length $K$ such that $\mathbb{P}\left[W \neq \hat{W}\right ] \leq \epsilon$ for any $\epsilon > 0$.
The capacity is the largest nonnegative rate that is $\epsilon$-achievable for $\epsilon \in (0,1)$.

The single-antenna static Gaussian HD diamond network $\mcal{N}_{[1:N]}$, shown in Fig.~\ref{fig:diamond2relays}, is defined by the input/output relationship\footnote{In the rest of the paper,  we drop the dependence of the channel inputs and outputs on the time $t$ in our expressions for ease of notation.}
\begin{subequations}
\label{eq:in_out_relation}
\begin{align}
     Y_{i} &= (1 - S_i) h_{is}X_0 + Z_i,\quad \forall i \in [1:N], \\
     Y_{N+1} &= \sum_{i=1}^N S_i h_{di}X_i + Z_{N+1}, 
\end{align}
\end{subequations}
where: 
(i) $S_i$ is the binary random variable that represents the state of the $i$-th relay, i.e., when $S_i = 0$ the $i$-th relay is receiving while when $S_i = 1$ the $i$-th relay is transmitting; 
(ii) $\left(h_{is},h_{di} \right )\in \mathbb{C}^2$ represent the channel coefficients from the source to the $i$-th relay and from the $i$-th relay to the destination, respectively; the channel gains are assumed to be constant for the whole transmission duration and hence known to all nodes;
(iii) the channel inputs are subject to a unitary average power constraint, i.e., $\mathbb{E} \left [|X_k|^2 \right ] \leq 1, k \in [0:N]$; 
(iv) $Z_i,$ $i \in [1:N+1]$ indicates the additive white Gaussian noise at the $i$-th node; noises are assumed to be independent and identically distributed as $\mcal{CN}(0,1)$.
We denote with $\ell_i$ and $r_i$ the individual link capacities, namely
\begin{subequations}
\label{eq:P2PLinkCap}
\begin{align}
\ell_i := \log\left( 1+ |h_{is}|^2\right),\quad \forall i \in [1:N], \\
r_i := \log\left( 1+ |h_{di}|^2\right), \quad \forall i \in [1:N].
\end{align}
\end{subequations}

The capacity of the Gaussian HD diamond network $\mcal{N}_{[1:N]}$ described in~\eqref{eq:in_out_relation} is not known in general, but from the works in~\cite{LimIT2011}, \cite{AvestimehrIT2011}, \cite{OzgurIT2013}, \cite{LimISIT2014}, it follows that it can be approximated to within a constant gap $G = O(N)$ by
\begin{align}
\label{eq:CapApp}
{\mathsf{C}}_{\mcal{N}_{[1:N]}} = \max_{\lambda \in \Lambda} \min_{\mcal{A} \subseteq [1:N]}\sum_{s \in [0:1]^N} \lambda_s 
\left( 
\max_{i \in \mcal{L}_s \cap \mathcal{A}} \ell_i+\max_{i \in  \mcal{T}_s \cap \mcal{A}^c} r_i  \right),
\end{align}
where: (i) $\Lambda = \{ \lambda : \lambda \in \mathbb{R}^{2^N},\ \lambda \geq 0,\ \|\lambda\|_1 = 1 \}$ is the set of all possible listen/transmit configuration states, with $\lambda_s = \mathbb{P} \left [S^N =s\right ] \in [0,1]$; 
(ii) $\mcal{L}_s$ (respectively, $\mathcal{T}_s$) represents the set of indices of relays listening (respectively, transmitting) in the relaying state $s \in [0:1]^N$, i.e., among the relays `on the side of the destination' (in~\eqref{eq:CapApp} indexed by $\mathcal{A}$) only those in receive mode matter, and similarly, among the relays `on the side of the source' (in~\eqref{eq:CapApp} indexed by $\mcal{A}^c =  [1:N]\backslash \mcal{A}$) only those in transmit mode matter.
For the particular case of $N=1$, the approximate capacity in~\eqref{eq:CapApp} becomes
\begin{align}
\label{eq:singleCapGen}
{\mathsf{C}}_{\mcal{N}_{\{1\}}} = \frac{\ell_1 r_1}{\ell_1 + r_1}
\end{align}
and when $N=2$ the authors in~\cite{BagheriIT2014} derived ${\mathsf{C}}_{\mcal{N}_{[1:2]}}$ in~\eqref{eq:CapApp} in closed form.

%
In what follows we say that the subnetwork $\mathcal{N}_{\mathcal{K}}$ with $\mathcal{K} \subseteq [1:N]$ operates with a `natural' schedule derived from the schedule $\lambda$ of $\mathcal{N}_{[1:N]}$ if the schedule of $\mathcal{N}_{\mathcal{K}}$ is constructed directly from $\lambda$, as better explained through the following example.

\noindent{\bf Example.} 
Consider a Gaussian HD diamond network $\mathcal{N}_{[1:N]}$ with $N=3$.
Let 
\begin{align*}
    \lambda = [\lambda_{000}\ \ \lambda_{001}\ \ \lambda_{010}\ \  \lambda_{011}\ \  \lambda_{100}\ \ \lambda_{101}\ \  \lambda_{110}\ \  \lambda_{111}]^T
\end{align*}
be a schedule for $\mathcal{N}_{[1:3]}$. 
Denote with $\lambda^{\left (\mcal{N}_{\{2,3\}} \right )}$ (respectively, $\lambda^{\left (\mcal{N}_{\{2\}} \right)}$) the schedule that is derived naturally from $\lambda$ for the subnetwork $\mcal{N}_{\{2,3\}}$ (respectively, $\mcal{N}_{\{2\}}$).
With this, we have
\begin{align*}
\lambda^{\left (\mcal{N}_{\{2,3\}} \right )}=[\lambda_{000}+\lambda_{100} \ \ \lambda_{001}+\lambda_{101} \ \ \lambda_{010}+\lambda_{110} \  \ \lambda_{011}+\lambda_{111}]^T
\end{align*}
and similarly we get
\begin{align*}
\lambda^{\left (\mcal{N}_{\{2\}} \right)} = [\lambda_{000}+\lambda_{001}+\lambda_{100}+\lambda_{101} \ \ \ \lambda_{010}+\lambda_{011}+\lambda_{110}+\lambda_{111}]^T.
\end{align*}
%
%
\noindent Thus, from the expression in~\eqref{eq:CapApp}, the approximate achievable rate ${\mathsf{R}}^{\lambda}_{\mathcal{N}_{\mathcal{K}}}$ of a subnetwork (for example $\mcal{N}_{\{2,3\}}$) when operating with the `natural' schedule derived from $\lambda$ is
\begin{align} 
\label{eq:subnetwork_using_full_schedule}
{\mathsf{R}}^{\lambda}_{\mathcal{N}_{\{2,3\}}} 
&= \min_{\mcal{A} \subseteq \{2,3\}} \sum_{s \in [0:1]^{2}} \lambda^{(\mcal{N}_{\{2,3\}})}_s \left( \max_{i \in \mathcal{A}} \ell_{i,s}^{\prime} +\max_{i \in \{2,3\} \backslash \mathcal{A}} r_{i,s}^{\prime} 
\right) \nonumber \\
&= \min_{\mcal{A} \subseteq \{2,3\}} \sum_{s \in [0:1]^3} \lambda_s 
\left(\max_{i \in \mathcal{A}} \ell_{i,s}^{\prime} +\max_{i \in \{2,3\} \backslash \mathcal{A}} r_{i,s}^{\prime} \right),
\end{align}
where
\begin{align}
\label{eq:NewChannels}
\ell_{i,s}^{\prime} = \left \{
\begin{array}{ll}
\ell_i & \text{if} \ i \in \mathcal{L}_s
\\
0 & \text{otherwise}
\end{array}
\right.,
\qquad 
r_{i,s}^{\prime} = \left \{
\begin{array}{ll}
r_i & \text{if} \ i \in \mathcal{T}_s
\\
0 & \text{otherwise}
\end{array}
\right. .
\end{align}

\section{Diamond Networks and Submodularity Properties}
\label{sec:Properties}
In this section we derive and discuss some properties of diamond networks and submodular functions, which represent the main ingredient in the proof of our main results.
It is worth noting that, beyond their utilization in the proofs, these properties might be of independent interest.
\subsection{A partition lemma for diamond networks}
The first result that we derive provides an upper bound on the approximate HD rate that can be achieved by the full network. This upper bound is stated  in the following lemma -- which we refer to as the {\it partition lemma} -- whose proof can be found in Appendix~\ref{app:lemmaPart}.

\begin{lem}[{\bf Partition lemma}]
\label{lem:part}
Let $\lambda$ be a schedule for the $N$-relay Gaussian HD diamond network $\mathcal{N}_{[1:N]}$. Then, for any $\mcal{K} \subseteq [1:N]$, we have
\begin{align}
\label{eq:PartEq}
{\mathsf{R}}^{\lambda}_{\mathcal{N}_{[1:N]}}
 \leq{\mathsf{R}}^{\lambda}_{\mathcal{N}_{\mathcal{K}}} +{\mathsf{R}}^{\lambda}_{\mathcal{N}_{[1:N]\backslash \mathcal{K}}},
\end{align}
where the subnetworks $\mathcal{N}_{\mathcal{K}}$ and $\mathcal{N}_{[1:N]\backslash \mathcal{K}}$ operate with the `natural' schedule derived from $\lambda$.
\end{lem}

The result in Lemma~\ref{lem:part} has the following two consequences:
\begin{enumerate}
\item Let $\lambda^\star$ be an optimal schedule for the full network $\mathcal{N}_{[1:N]}$, i.e., ${\mathsf{R}}^{\lambda^\star}_{\mathcal{N}_{[1:N]}} = {\mathsf{C}}_{\mathcal{N}_{[1:N]}}$.
Since the `natural' schedule constructed from $\lambda^\star$ might not be the optimal one for the subnetworks $\mathcal{N}_{\mathcal{K}}$ and $\mathcal{N}_{[1:N]\backslash \mathcal{K}}$, then ${\mathsf{R}}^{\lambda^\star}_{\mathcal{N}_{\mathcal{K}}} \leq {\mathsf{C}}_{\mathcal{N}_{\mathcal{K}}}$ and similarly ${\mathsf{R}}^{\lambda^\star}_{\mathcal{N}_{[1:N]\backslash \mathcal{K}}} \leq {\mathsf{C}}_{\mathcal{N}_{[1:N]\backslash \mathcal{K}}}$.
Hence, 
the result in Lemma~\ref{lem:part} straightforwardly implies that 
\begin{align}
\label{eq:newEqPart}
{\mathsf{C}}_{\mathcal{N}_{[1:N]}}
 \leq{\mathsf{C}}_{\mathcal{N}_{\mathcal{K}}}  +{\mathsf{C}}_{\mathcal{N}_{[1:N]\backslash \mathcal{K}}}.
\end{align}
For example, consider $\mathcal{N}_{[1:N]}$ with $N=3$. 
The inequality above implies that
\begin{align*}
&{\mathsf{C}}_{\mathcal{N}_{[1:3]}} \leq {\mathsf{C}}_{\mathcal{N}_{\{1,2\}}}+{\mathsf{C}}_{\mathcal{N}_{\{3\}}},\ \
{\mathsf{C}}_{\mathcal{N}_{[1:3]}} \leq {\mathsf{C}}_{\mathcal{N}_{\{1,3\}}}+{\mathsf{C}}_{\mathcal{N}_{\{2\}}},\\
&{\mathsf{C}}_{\mathcal{N}_{[1:3]}} \leq {\mathsf{C}}_{\mathcal{N}_{\{2,3\}}}+{\mathsf{C}}_{\mathcal{N}_{\{1\}}},\ \
{\mathsf{C}}_{\mathcal{N}_{[1:3]}} \leq \sum_{i=1}^3 {\mathsf{C}}_{\mathcal{N}_{\{i\}}}.
\end{align*}
\item The result in Lemma~\ref{lem:part} can be used to answer the following question: if we remove a link of capacity $\delta$ can we decrease the approximate capacity by more than $\delta$?
    This question was firstly formulated in the network coding domain~\cite{HoAllerton2010,JalaliITA2011} where the authors sought to understand whether removing a single edge of capacity $\delta$ can change the capacity region of the network by more than $\delta$ in each dimension.
This is an open problem in general and the question has been answered only for some particular cases.
The result in Lemma~\ref{lem:part} implies that, for Gaussian HD diamond networks\footnote{Thanks to the result in  Lemma~\ref{lemma:FDpart} in Appendix~\ref{app:lemmaPart}, the same statement also holds for Gaussian FD diamond networks.}, removing a link of capacity $\delta$ cannot decrease the approximate capacity by more than $\delta$. 
In fact, without loss of generality, let $\delta= \ell_i$, for some $i \in [1:N]$ (the same holds for $\delta=r_i$).
Then, from~\eqref{eq:newEqPart}, we have
\begin{align*}
\mathsf{C}_{\mathcal{N}_{[1:N]}} \leq \mathsf{C}_{{\bar{\mathcal{N}}}_i} + \mathsf{C}_{\mathcal{N}_{\{i\}}}
\implies
\mathsf{C}_{{\bar{\mathcal{N}}}_i} &\geq \mathsf{C}_{\mathcal{N}_{[1:N]}}-\mathsf{C}_{\mathcal{N}_{\{i\}}}
\\& 
\stackrel{{\rm{(a)}}}{\geq} \mathsf{C}_{\mathcal{N}_{[1:N]}} - \min \left \{ \delta, r_i\right \}
\geq \mathsf{C}_{\mathcal{N}_{[1:N]}} - \delta,
\end{align*}
where the inequality in $\rm{(a)}$ follows since $\mathsf{C}_{\mathcal{N}_{\{i\}}} \leq \mathsf{C}_{\mathcal{N}_{\{i\}}}^{\rm{FD}} = \min \left \{ \delta, r_i\right \}$.
\end{enumerate}

\subsection{Submodular functions and cut properties}
We now derive a property of submodular functions, which we next leverage to prove a property on cuts in diamond networks.

\begin{defin}
\label{def:subFunc}
For a finite set $\Omega$, let $f: 2^\Omega \to \mathbb{R}$ be a set function defined on $\Omega$. The set function $f$ is submodular if
\begin{align}
\forall \mcal{A}, \mcal{B} \subseteq \Omega,\quad f(\mcal{A}) + f(\mcal{B}) \geq f(\mcal{A} \cup \mcal{B}) + f(\mcal{A} \cap \mcal{B}).
\label{eq:submod_def}
\end{align}
\end{defin}
Building on the definion in~\eqref{eq:submod_def}, we now prove a property for a general submodular function.
\begin{lem}
\label{lemma:subProperty}
Let $f$ be a submodular set function defined on $\Omega$.
Then, for any group of $n$ sets $\mathcal{A}_i \subseteq \Omega$, $i \in [1:n]$,
\begin{align*}
\sum_{i=1}^n f \left( \mathcal{A}_i \right) \geq
\sum_{j=1}^n f \left( \mathcal{E}^{(n)}_j \right),
\end{align*}
where $\mathcal{E}_j^{(n)}$ is the set of elements that appear in at least $j$ sets $\mathcal{A}_i, i \in [1:n]$.
\end{lem}
\begin{IEEEproof}
The proof relies on the definition of submodular functions and on some set-theoretic properties. 
The detailed proof can be found in Appendix~\ref{appendix:proof_of_subproperty}.
\end{IEEEproof}
To better understand what Lemma~\ref{lemma:subProperty} implies, consider the following example.

\noindent {\bf Example.} Let $\Omega = [1:7]$ and consider the subsets $\mcal{A}_1 = \{1,2,5,7\},\ \mcal{A}_2 = \{4,5\},\ \mcal{A}_3 = \{2,4,5,6\}$.
Lemma~\ref{lemma:subProperty} proves that, for a submodular set function $f$ defined over $\Omega$, we get
\begin{align} 
\label{examp:subProperty}
f(&\overbrace{\{1,2,5,7\}}^{\mcal{A}_1}) + f(\overbrace{\{4,5\}}^{\mcal{A}_2}) + f(\overbrace{\{2,4,5,6\}}^{\mcal{A}_3})  \geq f(\underbrace{\{1,2,4,5,6,7\}}_{\mcal{E}^{(3)}_1}) + f(\underbrace{\{2,4,5\}}_{\mcal{E}^{(3)}_2}) + f(\underbrace{\{5\}}_{\mcal{E}^{(3)}_3}).
\end{align}
Now, as an example, consider $f(\mcal{A}) = \displaystyle\max_{i \in \mcal{A}}\{ i\}$ for $\mcal{A} \subseteq \Omega$, which is a submodular set function.
By evaluating both sides of~\eqref{examp:subProperty} for our example function, we get
\begin{align*}
\sum_{i=1}^3 f(\mcal{A}_i)  &=7 +5 + 6 = 18, \quad  \sum_{i=1}^3 f \left (\mcal{E}^{(3)}_j \right ) = 7 + 5 + 5 = \!17\\
&\implies \sum_{i=1}^3 f(\mcal{A}_i) \geq \sum_{i=1}^3 f \left (\mcal{E}^{(3)}_j \right)\ .
\end{align*}

Next, we use the result on submodular functions in Lemma~\ref{lemma:subProperty} to prove the following result for Gaussian diamond networks.
\begin{lem}
\label{Lemma:FD3From4}
Consider an $N$-relay Gaussian diamond network $\mathcal{N}_{[1:N]}$. 
Then, for any collection of sets $\mathcal{A}_i \subseteq [1:N]\backslash \{i\}$, there exists a collection of $(N-1)$ sets $\mathcal{A}_{{\rm{F}}j} \subseteq [1:N]$, with $j \in [1:N-1]$ such that
\begin{align}
\label{eq:forEx}
&\sum_{j=1}^N \left(\max_{i \in \mathcal{A}_j} \ell_i 
+\max_{i \in ([1:N]\backslash \{j\})\backslash \mathcal{A}_j} r_i \right)
 \geq \sum_{j=1}^{N-1}
\left(\max_{i \in \mathcal{A}_{{\rm{F}}j}} \ell_i 
+ \max_{i \in [1:N] \backslash \mathcal{A}_{{\rm{F}}j}} r_i \right).
\end{align}
Moreover the sets $\mcal{A}_{{\rm F}j}$ do not depend on the values $(\ell_i,r_i)$.
\end{lem}

\begin{IEEEproof}
    The proof, which is based on the result in Lemma~\ref{lemma:subProperty} and on simple counting arguments, can be found in Appendix~\ref{app:LemmaDiamond}.
\end{IEEEproof}

We next provide a simple example that better explains the implication of Lemma~\ref{Lemma:FD3From4}.
 
\noindent {\bf Example.}
Consider a $3$-relay Gaussian diamond network $\mathcal{N}_{[1:3]}$.
With this, we have $\bar{\mathcal{N}}_1 = \mcal{N}_{\{2,3\}}$, $\bar{\mathcal{N}}_2= \mcal{N}_{\{1,3\}}$ and $\bar{\mathcal{N}}_3 = \mcal{N}_{\{1,2\}}$.
Now for the subnetwork $\bar{\mathcal{N}}_i$ consider the following possible cut $\mathcal{A}_i$:
(i) $\mathcal{A}_1 = \emptyset$ (i.e., in $\bar{\mathcal{N}}_1$ relays $2$ and $3$ are `on the side of the source');
(ii) $\mathcal{A}_2 = \{3\}$ (i.e., in $\bar{\mathcal{N}}_2$ relay $1$ is `on the side of the source' and relay $3$ is `on the side of the destination');
(iii) (i) $\mathcal{A}_3 = \{1,2\}$ (i.e., in $\bar{\mathcal{N}}_3$ relays $1$ and $2$ are `on the side of the destination').
With this, by evaluating the left-hand side of~\eqref{eq:forEx}, we obtain
\begin{align*}
\sum_{j=1}^3 \left(\max_{i \in \mathcal{A}_j} \ell_i 
+\max_{i \in ([1:3]\backslash \{j\})\backslash \mathcal{A}_j} r_i \right) &= 
\max_{i \in \{2,3\}} r_i + \ell_3 +r_1 + \max_{i \in \{1,2\}} \ell_i
\\& \geq \max_{i \in [1:3]} \ell_i + \max_{i \in [1:3]} r_i
\\& = \sum_{j=1}^{2}
\left(\max_{i \in \mathcal{A}_{{\rm{F}}j}} \ell_i 
+ \max_{i \in [1:3] \backslash \mathcal{A}_{{\rm{F}}j}} r_i \right),
\end{align*}
where we let $\mathcal{A}_{{\rm{F}}1} = \emptyset$ and $\mathcal{A}_{{\rm{F}}2} = [1:3]$.
In this example, we considered a specific choice of $\mathcal{A}_i, i \in [1:3]$ in $\bar{\mathcal{N}}_i$.
By repeating the same reasoning, it is possible to show that, for any of the $2^{N(N-1)} = 4^3$ possible combinations of cuts $\mathcal{A}_i$, there always exist two cuts $\mathcal{A}_{{\rm{F}}j}, j \in [1:2]$ in the full network $\mathcal{N}_{[1:3]}$ such that~\eqref{eq:forEx} holds.

Before concluding this section and going into the technical details of how to use these results to prove our main results, we state a couple of remarks.

\begin{rem}
{\rm 
By considering the specific values of the link capacities ($\ell_i,r_i$) in a given network, we could prove the inequality in Lemma~\ref{Lemma:FD3From4} with a different construction than the one discussed in Appendix~\ref{app:LemmaDiamond}. 
The key property of the construction discussed in Appendix~\ref{app:LemmaDiamond} is that it is independent of ($\ell_i,r_i$). 
This becomes of fundamental importance when we consider HD cuts, as we will see in Section~\ref{sec:FundGar} when we prove Theorem~\ref{th:LowerBoundN-1OutOfN}.
}
\end{rem}

\begin{rem}
{\rm  
If the network and its subnetworks operate in FD, then Lemma~\ref{Lemma:FD3From4} directly relates cuts of the subnetworks $\bar{\mcal{N}}_i$ to cuts of the full network $\mcal{N}_{[1:N]}$ (see also the example above).
Furthermore, by choosing $\mcal{A}_i$ to be the minimum FD cut of the subnetwork $\bar{\mcal{N}}_i$, we get
\begin{align*}
    N \max_{i \in [1:N]} {\mathsf{C}}^{{\rm{FD}}}_{\bar{\mathcal{N}}_i} \geq \sum_{i=1}^N \mathsf{C}^{\rm FD}_{\bar{\mcal{N}}_i} &\geq  \sum_{j=1}^{N-1}
\left(\max_{i \in \mathcal{A}_{{\rm{F}}j}} \ell_i 
+ \max_{i \in [1:N] \backslash \mathcal{A}_{{\rm{F}}j}} r_i \right) \geq (N-1) \mathsf{C}^{\rm FD}_{[1:N]}.
\end{align*}
This is a different way of proving the result in~\cite[Theorem 1]{NazarogluIT2014} for $k=N-1$.}
\end{rem}

\section{A Simple Selection Algorithm}
\label{sec:MotivatingExample}
In this section, we investigate the performance (in terms of achievable fraction) of a simple algorithm that selects a subnetwork of $k=N-1$ relays.
In particular, the algorithm computes the $N$ single approximate capacities (see the expression in~\eqref{eq:singleCapGen}) and removes the worst relay, i.e., the one with the smallest single approximate capacity.
Since computing the single relay approximate capacities in a Gaussian HD diamond network with $N$ relays requires $O(N)$ operations, this algorithm runs in linear time and outputs an $(N-1)$-relay subnetwork whose performance guarantee is provided in the following theorem.


\begin{thm}
\label{thm:Removing_worst_k}
Consider a Gaussian HD diamond network $\mathcal{N}_{[1:N]}$.
Then, there always exists $i \in [1:N]$ such that
we can guarantee at least 
$\mathsf{C}_{{\bar{\mathcal{N}}}_i} \geq \frac{1}{2}{\mathsf{C}}_{\mathcal{N}_{[1:N]}}$. 
Moreover, if only the single relay approximate capacities are known\footnote{With this, we mean that the algorithm only leverages the expression of $\mathsf{C}_{\mathcal{N}_{\{i\}}}, \forall i \in [1:N]$ in~\eqref{eq:singleCapGen}, i.e., the algorithm is unaware of the values of the single link capacities $(\ell_i,r_i)$ in~\eqref{eq:P2PLinkCap}.}, then this bound is tight.
\end{thm}


\begin{IEEEproof}
We argue the lower bound in Theorem~\ref{thm:Removing_worst_k} by contradiction. 
Without loss of generality, let $\mathsf{C}_{\mathcal{N}_{\{N\}}} \leq \min_{i \in [1:N]} \mathsf{C}_{\mathcal{N}_{\{i\}}}$, i.e., the $N$-th relay is the worst.
Assume that $\mathsf{C}_{\mathcal{N}_{[1:N-1]}}< \frac{1}{2} \mathsf{C}_{\mathcal{N}_{[1:N]}}$. 
From the implication of Lemma~\ref{lem:part} in~\eqref{eq:newEqPart}, we have $\mathsf{C}_{\mathcal{N}_{[1:N-1]}} + \mathsf{C}_{\mathcal{N}_{\{N\}}} \geq \mathsf{C}_{\mathcal{N}_{[1:N]}}$.
This property, together with the assumption that $\mathsf{C}_{\mathcal{N}_{[1:N-1]}}< \frac{1}{2} \mathsf{C}_{\mathcal{N}_{[1:N]}}$, implies that $\mathsf{C}_{\mathcal{N}_{\{N\}}} \geq \frac{1}{2}\mathsf{C}_{\mathcal{N}_{[1:N]}}$.
However, since the relay number $N$ has the lowest approximate HD capacity, then $\forall j \in {[1:N-1]},\ \mathsf{C}_{\mcal{N}_{\{j\}}} \geq \frac{1}{2}\mathsf{C}_{\mathcal{N}_{[1:N]}}$.
Therefore, we finally have the following contradiction
\begin{align*}
    \forall j \in {[1:N-1]},\quad 
\frac{1}{2}\mathsf{C}_{\mathcal{N}_{[1:N]}} \leq \mathsf{C}_{\mcal{N}_{\{j\}}} \leq \mathsf{C}_{\mathcal{N}_{[1:N-1]}}< \frac{1}{2} \mathsf{C}_{\mathcal{N}_{[1:N]}}.
\end{align*}
%
This concludes the proof of the lower bound in Theorem~\ref{thm:Removing_worst_k}.

To prove that the bound in Theorem~\ref{thm:Removing_worst_k} is indeed tight it suffices to provide a network construction where having the knowledge of only the single relay approximate capacities does not guarantee that a subnetwork $\mathcal{N}_{\mcal{K}}$ of $N-1$ relays, with $\mathsf{C}_{\mathcal{N}_{\mcal{K}}}$
strictly greater than 
$\frac{1}{2}{\mathsf{C}}_{\mathcal{N}_{[1:N]}}$, can be chosen deterministically. 
For $N \geq 2$, let
\begin{subequations}
\label{eq:NetExHalf}
\begin{align}
&\ell_i = \frac{1}{2},\quad r_i = L,  \qquad \forall i \in [1:N-1],\\
&\ell_N = L, \quad r_N = \frac{1}{2},
\end{align}
\end{subequations}
where $L \rightarrow \infty$.
Note that for the network construction in~\eqref{eq:NetExHalf} we have:
(i) $\forall i \in [1:N],\ \mathsf{C}_{\mathcal{N}_{\{i\}}} = \frac{1}{2}$ and
(ii) the approximate HD capacity of the full network is ${\mathsf{C}}_{\mathcal{N}_{[1:N]}}=1$.
We now want to remove the worst relay based only on the knowledge of the single relay approximate capacities.
Since these are all equal, then one can choose to remove one relay uniformly at random.
If the $N$-th relay is removed, then the remaining network has an approximate capacity of 
$\mathsf{C}_{\mathcal{N}_{[1:N-1]}} = \frac{1}{2}$,
which shows that the lower bound in Theorem~\ref{thm:Removing_worst_k} is indeed tight if the choice (of which relay to remove) is based only on the single relay approximate capacities.
\end{IEEEproof}

The tightness argument in Theorem~\ref{thm:Removing_worst_k} implies that, for an algorithm that removes the worst relay - by only computing the single relay approximate capacities - no higher worst-case guarantee can be provided.
However, this result is pretty conservative.
In fact, with reference to the specific network construction in~\eqref{eq:NetExHalf}, if we are allowed to select $N-1$ relays based on the approximate capacities of the $2$-relay subnetworks, then we would never remove the $N$-th relay. 
This is because any $2$-relay subnetwork which involves the $N$-th relay has an approximate capacity of 
$\mathsf{C}_{\mathcal{N}_{\{N,i\}}}=1=\mathsf{C}_{\mathcal{N}_{[1:N]}}, \forall i \in [1:N-1]$.
This simple example suggests that a smarter choice (compared to the one based on removing the worst relay) of which $N-1$ relays to select might lead to a higher worst-case achievable fraction, compared to the $\frac{1}{2}$ in Theorem~\ref{thm:Removing_worst_k}.
In the next section, we will formally prove that this observation is indeed true.
Before concluding this section, we next generalize the lower bound in Theorem~\ref{thm:Removing_worst_k} to generic values of $k \in [1:N]$.

\subsection{The general case $k \in [1:N]$}
We now generalize the lower bound in Theorem~\ref{thm:Removing_worst_k} when $k \in [1:N]$.
Towards this end, we consider an algorithm that removes the worst $N-k$ relays (i.e., those with the lowest single relay approximate capacities) from the network of $N$ relays.
The algorithm first computes the single relay approximate capacities -- which requires $O(N)$ operations.
It then orders the relays in descending order based on their single approximate capacities, i.e., in this new ordering the first relay is the one for which $\mathsf{C}_{\mathcal{N}_{\{1\}}} \geq \max_{i \in [2:N]} \mathsf{C}_{\mathcal{N}_{\{i\}}}$, the second relay is the one for which $\mathsf{C}_{\mathcal{N}_{\{2\}}} \geq \max_{i \in [3:N]} \mathsf{C}_{\mathcal{N}_{\{i\}}}$ and so on till the $N$-th relay for which $\mathsf{C}_{\mathcal{N}_{\{N\}}} = \min_{i \in [1:N]} \mathsf{C}_{\mathcal{N}_{\{i\}}}$; this step requires $O(N\log(N))$ operations.
Finally, the algorithm discards the last $N-k$ relays.
In other words, the algorithm runs in $O(N\log(N))$ and outputs a $k$-relay subnetwork whose performance guarantee is provided in the following lemma.
\begin{lem}
\label{lem:genKsimple}
Consider a Gaussian HD diamond network $\mathcal{N}_{[1:N]}$ where the relays are ordered in descending order based on their single approximate capacities.
By operating only the relays in $\mathcal{N}_{[1:k]}$, we can always guarantee at least $\mathsf{C}_{\mathcal{N}_{[1:k]}} \geq 2^{-(N-k)}{\mathsf{C}}_{\mathcal{N}_{[1:N]}}$. 
\end{lem}

\begin{IEEEproof}
Clearly, for the case $k=N-1$ the lower bound in Lemma~\ref{lem:genKsimple} is equivalent to the one in Theorem~\ref{thm:Removing_worst_k}.
We now argue the lower bound in Lemma~\ref{lem:genKsimple} by contradiction. 
Without loss of generality, assume that instead of removing the last $N-k$ relays all together (recall that relays are ordered in descending order based on their single approximate capacities), we remove them in $N-k$ steps, i.e., at step $i \in [1:N-k]$ we remove the relay number $N-i+1$.
Assume that at step $i$ we have that $\mathsf{C}_{\mathcal{N}_{[1:N-i]}}< \frac{1}{2} \mathsf{C}_{\mathcal{N}_{[1:N-i+1]}}$. 
From~\eqref{eq:newEqPart}, we have $\mathsf{C}_{\mathcal{N}_{[1:N-i]}} + \mathsf{C}_{\mathcal{N}_{\{N-i+1\}}} \geq \mathsf{C}_{\mathcal{N}_{[1:N-i+1]}}$.
This property, together with the assumption that $\mathsf{C}_{\mathcal{N}_{[1:N-i]}}< \frac{1}{2} \mathsf{C}_{\mathcal{N}_{[1:N-i+1]}}$, implies that $\mathsf{C}_{\mathcal{N}_{\{N-i+1\}}} \geq \frac{1}{2}\mathsf{C}_{\mathcal{N}_{[1:N-i+1]}}$.
However, since the relay number $N-i+1$ has the lowest approximate HD capacity at step $i$, then $\forall j \in {[1:N-i]},\ \mathsf{C}_{\mcal{N}_{\{j\}}} \geq \frac{1}{2}\mathsf{C}_{\mathcal{N}_{[1:N-i+1]}}$.
Therefore, we finally have the following contradiction
\begin{align*}
    \forall j \in {[1:N-i]},\quad 
\frac{1}{2}\mathsf{C}_{\mathcal{N}_{[1:N-i+1]}} \leq \mathsf{C}_{\mcal{N}_{\{j\}}} \leq \mathsf{C}_{\mathcal{N}_{[1:N-i]}}< \frac{1}{2} \mathsf{C}_{\mathcal{N}_{[1:N-i+1]}}.
\end{align*}
Thus, $\forall i \in [1:N-k]$, we have that $\mathsf{C}_{\mathcal{N}_{[1:N-i]}} \geq  \frac{1}{2} \mathsf{C}_{\mathcal{N}_{[1:N-i+1]}}$.
By recursively applying this expression $(N-k)$ times we are left with a $k$-relay subnetwork that achieves an approximate capacity $\mathsf{C}_{\mathcal{N}_{[1:k]}} \geq 2^{-(N-k)}{\mathsf{C}}_{\mathcal{N}_{[1:N]}}$.
This concludes the proof.
\end{IEEEproof}

\section{A Fundamental Guarantee for Selecting $N-1$ Relays}
\label{sec:FundGar}

In this section we derive a fundamental guarantee (in terms of achievable fraction) when $N-1$ relays are selected out of the $N$ possible ones.
We assert that this guarantee is fundamental because it represents the highest worst-case fraction that can be guaranteed when $N-1$ relays are selected, independently of the actual values of the channel parameters.
In particular, our main result is stated in the following theorem.

\begin{thm}
\label{th:LowerBoundN-1OutOfN}
For any $N$-relay Gaussian HD diamond network $\mathcal{N}_{[1:N]}$, there always exists a subnetwork $\mathcal{N}_{\mathcal{K}}$, with $|\mathcal{K}|=N-1$, that achieves at least $\mathsf{C}_{\mathcal{N}_{\mathcal{K}}} \geq \frac{N-1}{N} {\mathsf{C}}_{\mathcal{N}_{[1:N]}}$.
Moreover, this bound is tight.
\end{thm}

\begin{IEEEproof}
In order to derive the lower bound in Theorem~\ref{th:LowerBoundN-1OutOfN}, we first state the following lemma, whose proof is based on Lemma~\ref{Lemma:FD3From4} and is delegated to Appendix~\ref{app:proof_mainresults}.
\begin{lem}
\label{Lemma:HD3From4}
Consider an arbitrary $N$-relay Gaussian HD diamond network $\mathcal{N}_{[1:N]}$ operated with the schedule $\lambda$. 
Then,
\begin{align}
\sum_{i=1}^N {\mathsf{R}}^{\lambda}_{{\bar{\mathcal{N}}}_i} \geq (N-1)\mathsf{R}^{\lambda}_{\mathcal{N}_{[1:N]}}.
\end{align}
\end{lem}
%
The lower bound in Theorem~\ref{th:LowerBoundN-1OutOfN} is a direct consequence of Lemma~\ref{Lemma:HD3From4} as explained in what follows.
Let $\lambda^\star$ be an optimal schedule for the full network $\mathcal{N}_{[1:N]}$, i.e., ${\mathsf{R}}^{\lambda^\star}_{\mathcal{N}_{[1:N]}} = {\mathsf{C}}_{\mathcal{N}_{[1:N]}}$.
Since the `natural' schedule constructed from $\lambda^\star$ might not be the optimal one for the subnetwork ${\bar{\mathcal{N}}}_i$, then clearly we have $\mathsf{R}^{\lambda^\star}_{{\bar{\mathcal{N}}}_i} \leq \mathsf{C}_{{\bar{\mathcal{N}}}_i}, \forall i \in [1:N]$.
Using the result in Lemma~\ref{Lemma:HD3From4} with $\lambda^\star$, we get
\begin{align*}
(N-1)\mathsf{C}_{\mathcal{N}_{[1:N]}} 
\leq 
\sum_{i=1}^N {\mathsf{R}}^{\lambda^\star}_{\bar{\mathcal{N}}_i} &
\leq \sum_{i=1}^N {\mathsf{C}}_{\bar{\mathcal{N}}_i} \leq N \max_{i\in [1:N]}  {\mathsf{C}}_{\bar{\mathcal{N}}_i}.
\end{align*}
Let $i^\star = \arg \max \left \{ {\mathsf{C}}_{\bar{\mathcal{N}}_i}  \right \}$. Then, by setting $\mathcal{K} = [1:N] \backslash \{i^\star\}$, we have that
\begin{align*}
\mathsf{C}_{\mathcal{N}_{\mathcal{K}}} \geq \frac{N-1}{N} {\mathsf{C}}_{\mathcal{N}_{[1:N]}}.
\end{align*}
This completes the proof of the lower bound in Theorem~\ref{th:LowerBoundN-1OutOfN}.

To prove that the ratio in Theorem~\ref{th:LowerBoundN-1OutOfN} is tight, it suffices to provide an example of an $N$-relay network where the best (i.e., the one with the largest approximate capacity) subnetwork of $N-1$ relays achieves an approximate capacity, which is exactly the fraction of the full network approximate capacity in Theorem~\ref{th:LowerBoundN-1OutOfN}. 
To this end, consider the following structure:
\begin{subequations}
\label{eq:WorstNetwEx}
\begin{align}
&\ell_{i} = \ell_{\left \lfloor \frac{N}{2} \right \rfloor +i} = \frac{2i}{N}, \ i \in  \left [1: \left \lfloor \frac{N}{2} \right \rfloor \right],
\\& r_{i} = r_{\left \lfloor \frac{N}{2} \right \rfloor +i}= \frac{N-2i+2}{N},  \ i \in  \left [1:\left \lfloor \frac{N}{2} \right \rfloor \right],
\\& \text{if }N\text{ is odd:} \ \ \ell_N = L, \ r_N = \frac{1}{N},
\end{align}
\end{subequations}
where $L \rightarrow \infty$.
\begin{figure}
\centering
\includegraphics[width=0.57\columnwidth]{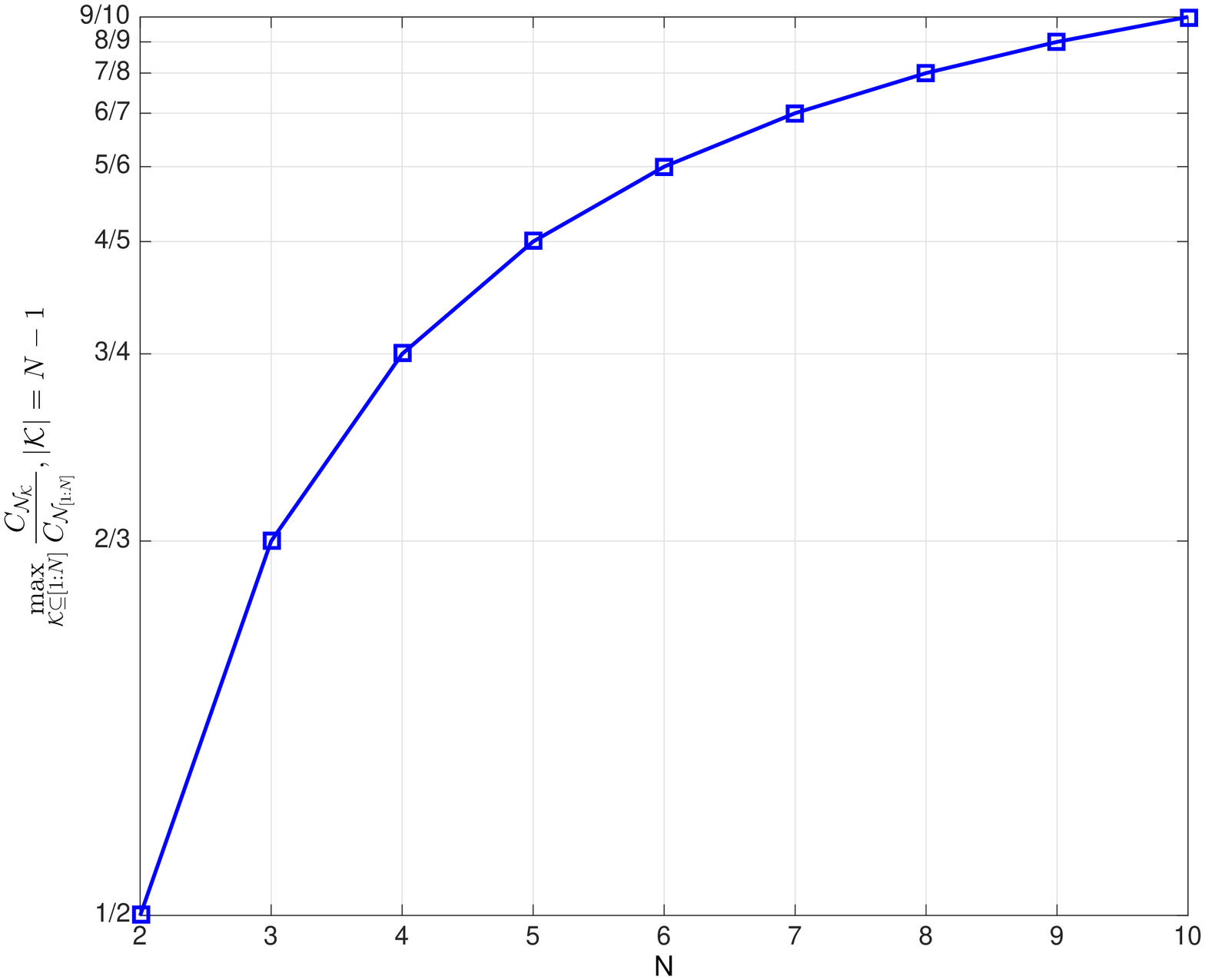}
\caption{$\displaystyle\max_{\mcal{K} \subseteq [1:N]}\frac{\mathsf{C}_{\mathcal{N}_{\mathcal{K}}}}{{\mathsf{C}}_{\mathcal{N}_{[1:N]}}}$ with $|\mathcal{K}|=N-1$ for the network in~\eqref{eq:WorstNetwEx} for $N \in [2:10]$.}
\label{fig:netwEval}
\end{figure}
Fig.~\ref{fig:netwEval} gives a representation of $\displaystyle\max_{\mcal{K} \subseteq [1:N]}\frac{\mathsf{C}_{\mathcal{N}_{\mathcal{K}}}}{{\mathsf{C}}_{\mathcal{N}_{[1:N]}}}$
for $N \in [2:10]$ with $|\mathcal{K}|=N-1$.
From Fig.~\ref{fig:netwEval} we observe that $\displaystyle\max_{\mcal{K} \subseteq [1:N]}\frac{\mathsf{C}_{\mathcal{N}_{\mathcal{K}}}}{{\mathsf{C}}_{\mathcal{N}_{[1:N]}}}=\frac{N-1}{N}$.
This completes the proof.
\end{IEEEproof}

Before concluding this section, we highlight some results, which are direct consequences of Lemma~\ref{Lemma:HD3From4} and Theorem~\ref{th:LowerBoundN-1OutOfN}.

\begin{rem}
{\rm
Theorem~\ref{th:LowerBoundN-1OutOfN} provides a performance guarantee that significantly improves over the one in Theorem~\ref{thm:Removing_worst_k}.
In fact, for high values of $N$, Theorem~\ref{th:LowerBoundN-1OutOfN} ensures that we can approach ${\mathsf{C}}_{\mathcal{N}_{[1:N]}}$ by operating only $N-1$ relays, which is twice the guarantee of $\frac{1}{2}{\mathsf{C}}_{\mathcal{N}_{[1:N]}}$ (independent of the value of $N$) provided by Theorem~\ref{thm:Removing_worst_k}.
}
\end{rem}


\begin{rem}
{\rm
The result in Theorem~\ref{th:LowerBoundN-1OutOfN} implies that, for any $N$-relay Gaussian HD diamond network, smartly removing one relay can reduce the approximate HD capacity of the network by at most $\frac{1}{N}$ of the full network approximate capacity.
We also highlight that the removed relay may not be the worst relay since in this case, as proved in Theorem~\ref{thm:Removing_worst_k}, we can guarantee only $\mathsf{C}_{\bar{\mathcal{N}}_{i}} \geq \frac{1}{2} {\mathsf{C}}_{\mathcal{N}_{[1:N]}}$, where $i \in [1:N]$ is the index of the worst relay.
However, for the specific network in~\eqref{eq:WorstNetwEx} the full network has an approximate capacity of 
${\mathsf{C}}_{\mathcal{N}_{[1:N]}}=1$
(see Appendix~\ref{app:cut1} for the detailed computation) and all the $(N-1)$-relay subnetworks have an approximate capacity of 
${\mathsf{C}}_{\mathcal{N}_{\mathcal{K}}}= \frac{N-1}{N}, \ \forall \mathcal{\mathcal{K}} \subseteq [1:N], \ |\mathcal{K}|=N-1$.
Hence, for this particular network, by removing {\it any} of the relays (i.e., the best or the worst), we always retain $\frac{N-1}{N}$ of the approximate capacity of the full network.
}
\end{rem}

\begin{cor} 
\label{corollary:complexity} 
Let $\lambda^\star$ be an optimal schedule of the full network $\mathcal{N}_{[1:N]}$, then:
\begin{enumerate}
\item For any $N$-relay Gaussian HD diamond network, there exists a subnetwork $\mcal{N}_{\mathcal{K}}$, with $|\mathcal{K}| = N-1$, such that, when operated with $\lambda^\star$, it satisfies that
\begin{align*}
{\mathsf{R}}^{\lambda^\star}_{\mathcal{N}_{\mathcal{K}}} \geq \frac{N-1}{N} \mathsf{C}_{\mathcal{N}_{[1:N]}}.
\end{align*}
\item There exist $N$-relay Gaussian HD diamond networks where $\lambda^\star$ can be used to naturally construct the optimal schedule for each subnetwork of $N-1$ relays (see for example, the network in~\eqref{eq:WorstNetwEx}).
\end{enumerate}
\end{cor}
\begin{rem}
{\rm
Corollary~\ref{corollary:complexity} implies that, to select a subnetwork of $N-1$ relays that guarantees the performance in Theorem~\ref{th:LowerBoundN-1OutOfN}, it is sufficient to know an optimal schedule $\lambda^\star$ of the whole network $\mathcal{N}_{[1:N]}$. 
In other words, by knowing $\lambda^\star$, there is no need to compute the optimal schedules for each of the $N$ subnetworks. 
This implies that, if $\lambda^\star$ can be used to construct a `natural' schedule for all $\mathcal{N}_{\mathcal{K}}$, with $|\mathcal{K}|=N-1$, in polynomial time, then a subnetwork $\mathcal{N}_{\mathcal{K}}$ that achieves the guarantee in Theorem~\ref{th:LowerBoundN-1OutOfN} can be discovered in polynomial time.
}
\end{rem}

We next leverage the result in Theorem~\ref{th:LowerBoundN-1OutOfN} to derive a lower bound for generic $k \in [1:N]$.

\subsection{The general case $k \in [1:N]$}

In this subsection we generalize the lower bound derived in Theorem~\ref{th:LowerBoundN-1OutOfN} when $k \in [1:N]$.
In particular, our result is stated in the following lemma.
\begin{lem}
\label{lem:LowerBoundkOutOfN}
Consider an arbitrary $N$-relay Gaussian HD diamond network $\mathcal{N}_{[1:N]}$ operated with the schedule $\lambda$. 
There always exists a subnetwork $\mathcal{N}_{\mathcal{K}}$ with $|\mathcal{K}|=k \in [1:N]$ that, when operated with the `natural' schedule derived from $\lambda$, achieves an approximate rate $\mathsf{R}^{\lambda}_{\mathcal{N}_{\mathcal{K}}}$
such that $\mathsf{R}^{\lambda}_{\mathcal{N}_{\mathcal{K}}} \geq \frac{k}{N} {\mathsf{R}}^{\lambda}_{\mathcal{N}_{[1:N]}}$.
\end{lem}

\begin{IEEEproof}
We recursively apply the result in Lemma~\ref{Lemma:HD3From4}.
We again let $\lambda$ be a schedule (not necessarily optimal) of the full $N$-relay network $\mathcal{N}_{[1:N]}$. 
With this we obtain
\begin{subequations}
\begin{align}
&\exists \ i_1 \in [1:N]  \ \text{such that for } \mcal{S}^{(1)} = \mathcal{N}_{[1:N]\backslash \{i_1\}}: \\
&\qquad {\mathsf{R}}^{\lambda}_{\mcal{S}^{(1)}} \geq \frac{N-1}{N} {\mathsf{R}}^{\lambda}_{\mathcal{N}_{[1:N]}}, \label{eq:it1lb}
\\
& \exists \ i_2 \in \mcal{S}^{(1)} =  \mathcal{N}_{[1:N]\backslash \{i_1\}}  \ \text{such that for } \mcal{S}^{(2)} = \mathcal{N}_{[1:N]\backslash \{i_{[1:2]}\}}: \nonumber \\
&\qquad 
{\mathsf{R}}^{\lambda}_{\mcal{S}^{(2)}}  \geq \frac{N-2}{N-1} {\mathsf{R}}^{\lambda}_{\mcal{S}^{(1)}}
\stackrel{\eqref{eq:it1lb}}{\geq} \frac{N-2}{N} {\mathsf{R}}^{\lambda}_{\mathcal{N}_{[1:N]}}, \label{eq:it2lb}
\\
& \exists \ i_3 \in \mcal{S}^{(2)} =  \mathcal{N}_{[1:N]\backslash \{i_{[1:2]}\}}   \ \text{such that for } \mcal{S}^{(3)} =  \mathcal{N}_{[1:N]\backslash \{i_{[1:3]}\}}: \nonumber \\
&\qquad 
{\mathsf{R}}^{\lambda}_{\mcal{S}^{(3)}} \geq \frac{N-3}{N-2} {\mathsf{R}}^{\lambda}_{\mcal{S}^{(2)}}
\stackrel{\eqref{eq:it2lb}}{\geq} \frac{N-3}{N} {\mathsf{R}}^{\lambda}_{\mathcal{N}_{[1:N]}},
\\&\qquad \qquad \qquad \qquad \qquad \vdots \nonumber
\\
& \exists \ i_{N-k} \in \mcal{S}^{(N-k-1)}   \ \text{such that for } \mcal{S}^{(N-k)} =  \mathcal{N}_{[1:N]\backslash \{i_{[1:N-k]}\}}: \nonumber \\
&\qquad {\mathsf{R}}^{\lambda}_{\mcal{S}^{(N-k)}} \geq \frac{k}{k+1} {\mathsf{R}}^{\lambda}_{\mcal{S}^{(N-k-1)}}
{\geq} \frac{k}{N} {\mathsf{R}}^{\lambda }_{\mathcal{N}_{[1:N]}},
\end{align}
\end{subequations}
which, since $\mathcal{S}^{(N-k)}$ contains $k$ relays, completes the proof.
\end{IEEEproof}

{\begin{rem}{\rm
Let $\lambda^\star$ be an optimal schedule for the full network $\mathcal{N}_{[1:N]}$, i.e., ${\mathsf{R}}^{\lambda^\star}_{\mathcal{N}_{[1:N]}} = {\mathsf{C}}_{\mathcal{N}_{[1:N]}}$.
Since the `natural' schedule constructed from $\lambda^\star$ might not be the optimal one for the subnetwork $\mathcal{N}_{\mathcal{K}}$, i.e., $\mathsf{R}^{\lambda^\star}_{\mathcal{N}_{\mathcal{K}}} \leq \mathsf{C_{\mathcal{N}_{\mathcal{K}}}}$, then
Lemma~\ref{lem:LowerBoundkOutOfN} provides a different bound from the one in~\cite{BrahmaISIT2014Relay} and from the $\frac{k}{2(k+1)}$ that is readily obtained from the result in~\cite{NazarogluIT2014}.
These bounds can be combined as
\begin{align} 
\label{eq:combined_bounds}
\frac{\mathsf{C_{\mathcal{N}_{\mathcal{K}}}}}{{\mathsf{C}}_{\mathcal{N}_{[1:N]}}}
\geq 
\begin{cases}
\max \left\{ \frac{1}{N}\ , \frac{1}{4} \right \}, & k=1 \\
\max \left\{ \frac{k}{N}\ , \frac{1}{2} \right \}, & N \geq k \geq 2 \\
\end{cases}.
\end{align}
From~\eqref{eq:combined_bounds}, we can see that in some cases (particularly when $k > N/2$), the new bound in Lemma~\ref{lem:LowerBoundkOutOfN} gives a better guarantee than those available in the literature.
Clearly, when $k=N-1$ the lower bound in~\eqref{eq:combined_bounds} is equivalent to the one in Theorem~\ref{th:LowerBoundN-1OutOfN}.
However, the lower bound in Lemma~\ref{lem:LowerBoundkOutOfN} is not tight for general $k \in [1:N]$.
Deriving tighter lower bounds is an interesting open problem, which is object of current investigation. 
For instance, for the case $k=1$, numerically we could not find network examples for which the fraction guarantee is less than $\frac{N}{4(N-1)}$.
}\end{rem}}

\begin{rem}
{\rm
The proof of Lemma~\ref{lem:LowerBoundkOutOfN} provides the blueprint for an algorithm that selects a subnetwork of $k$ relays that achieves the guarantee in the lemma.
The algorithm operates iteratively as follows. 
On the first iteration, given a network $ \mcal{N}^{(0)}= \mathcal{N}_{[1:N]}$ with $N$ relays and an operating schedule $\lambda$, we find a subnetwork $\mcal{N}^{(1)}$ with $N-1$ relays such that $\mcal{N}^{(1)}$, when operated with the `natural' schedule derived from $\lambda$, satisfies the bound in Lemma~\ref{lem:LowerBoundkOutOfN} for $k=N-1$. 
We can repeat the previous iteration $(N-k)$ times where on iteration $i$, we remove one relay to select a subnetwork $\mcal{N}^{(i)}$ such that 
\begin{align*}
 \mathsf{R}^{\lambda}_{\mcal{N}^{(i)}}\geq \frac{N-i}{N-i+1}    \mathsf{R}^{\lambda}_{\mcal{N}^{(i-1)}}.
\end{align*}
It is clear that after $(N-k)$ iterations, we have a subnetwork $\mcal{N}^{(N-k)}$ that contains exactly $k$ relays and for which
\begin{align*}
\mathsf{R}^{\lambda}_{\mcal{N}^{(N-k)}}\geq \frac{k}{N}    \mathsf{R}^{\lambda}_{\mcal{N}_{[1:N]}}.
\end{align*}
In~\cite{EtkinIT2014} the authors showed that the problem of computing the approximate capacity of a Gaussian FD relay network can be cast as a minimization problem of a submodular function, which can be solved in polynomial time. 
Therefore, if the fixed schedule $\lambda$ at which $\mathcal{N}_{[1:N]}$ is operated can be used to construct a `natural' schedule for $\mcal{N}^{(1)}$ in polynomial time, then the algorithm described above runs in polynomial time and provides the fraction guarantee in Lemma~\ref{lem:LowerBoundkOutOfN}.
}\end{rem}

\section{Discussion and Conclusions}
\label{sec:Concl}
In this section, we discuss some implications of the results derived in the previous sections and highlight differences between the selection performances in HD and FD diamond networks. 
We believe that the reason for this different behavior is that in HD the schedule plays a key role, i.e., removing some of the relays can change the optimal schedule of the remaining network.

\smallskip
\noindent {\bf{1) In HD the guarantee on  
            $\displaystyle\max_{\mcal{K} \subseteq [1:N]}\frac{\mathsf{C}_{\mathcal{N}_{\mathcal{K}}}}{{\mathsf{C}}_{\mathcal{N}_{[1:N]}}}$ for $|\mathcal{K}|=k\in [1:2]$}
decreases as $N$ increases.}
We here show that in HD, for the case $|\mathcal{K}|=k\in [1:2]$, the worst case fraction $\displaystyle\max_{\mcal{K} \subseteq [1:N]}\frac{\mathsf{C}_{\mathcal{N}_{\mathcal{K}}}}{{\mathsf{C}}_{\mathcal{N}_{[1:N]}}}$ depends on $N$ and decreases as $N$ increases.
This represents a surprising difference with respect to FD (where the worst case ratio for a fixed value of $k$ does not depend on $N$) and shows that FD and HD relay networks have a different nature.
In particular, from the result in Theorem~\ref{th:LowerBoundN-1OutOfN} for $|\mathcal{K}|=k\in [1:2]$ and $N = k+1$, we have $\frac{\mathsf{C}_{\mathcal{N}_{\mathcal{K}}} }{{\mathsf{C}}_{\mathcal{N}_{[1:N]}}} \geq \frac{k}{k+1}$ as in FD~\cite[Theorem 1]{NazarogluIT2014}.
However, in the regime $N \gg 1$, these values reduce to $\frac{\mathsf{C}_{\mathcal{N}_{\mathcal{K}}} }{{\mathsf{C}}_{\mathcal{N}_{[1:N]}}} \geq  \frac{1}{4}$ for $k=1$ and to $\frac{\mathsf{C}_{\mathcal{N}_{\mathcal{K}}} }{{\mathsf{C}}_{\mathcal{N}_{[1:N]}}} \geq \frac{1}{2}$ for $k=2$. 
Notice that these values coincide with the lower bounds:
(i) of $\frac{k}{2(k+1)}$ for $k=1$, which is readily obtained from the result in~\cite{NazarogluIT2014} by letting the selected relay listen for half of the time and transmit for the other half of the time;
(ii) derived in~\cite{BrahmaISIT2014Relay} for the case $k=2$, where the $2$ selected relays operate in a complementary fashion.
In particular, we have
\begin{thm}
\label{thm:highN}
There exist Gaussian HD diamond networks for which, when $N \gg 1$, the best subnetwork $\mcal{N}_{\mcal{K}}$ gives
\begin{align} 
\label{eq:highN}
\frac{\mathsf{C}_{\mathcal{N}_{\mathcal{K}}} }{{\mathsf{C}}_{\mathcal{N}_{[1:N]}}} =
\begin{cases}
\frac{1}{4}, & |\mathcal{K}|=1 \\
\frac{1}{2} , & |\mathcal{K}|=2 \\
\end{cases}.
\end{align}
\end{thm}

\begin{IEEEproof}
Consider the network in~\eqref{eq:WorstNetwEx}.
The best subnetwork $\mathcal{N}_{\mathcal{K}}$ with $|\mathcal{K}|=1$ achieves
\begin{align*}
\frac{\mathsf{C}_{\mathcal{N}_{\mathcal{K}}} }{{\mathsf{C}}_{\mathcal{N}_{[1:N]}}} = \frac{N+2}{4N},
\end{align*}
which for $N \gg 1$ gives $\frac{\mathsf{C}_{\mathcal{N}_{\mathcal{K}}} }{{\mathsf{C}}_{\mathcal{N}_{[1:N]}}} = \frac{1}{4}$,
while the best subnetwork $\mathcal{N}_{\mathcal{K}}$ with $|\mathcal{K}|=2$ relays achieves
\begin{align*}
\frac{\mathsf{C}_{\mathcal{N}_{\mathcal{K}}} }{{\mathsf{C}}_{\mathcal{N}_{[1:N]}}} = \frac{N+2}{2N},
\end{align*}
which for $N \gg 1$ gives $\frac{\mathsf{C}_{\mathcal{N}_{\mathcal{K}}} }{{\mathsf{C}}_{\mathcal{N}_{[1:N]}}} = \frac{1}{2}$.
We refer the reader to Appendix~\ref{app:cut1} for a detailed computation of these values.
\end{IEEEproof}

\smallskip 
\noindent{\bf{2) The best HD and FD subnetworks are not necessarily the same.}}
We next provide a couple of examples where we show that the best relay in HD and in FD might not be necessarily the same.
As a first example, consider a Gaussian $2$-relay diamond network with $\ell_1=1$, $\ell_2=\frac{2}{5}$, $r_1=\frac{1}{2}$ and $r_2=\frac{14}{5}$.
It is not difficult to see that if the relays operate in FD, then the first relay is the best and it achieves $ \mathsf{C}^{{\rm{FD}}}_{\mcal{N}_{\{1\}}} =\frac{1}{2}$, while if the relays operate in HD then the second relay is the best giving 
$\mathsf{C}_{\mcal{N}_{\{2\}}} =\frac{7}{20}$ (see the expression in~\eqref{eq:singleCapGen}).
As a second example consider a Gaussian $3$-relay diamond network with $\ell_{[1:3]}=r_{[1:2]}= \ell >0$ and $r_3 = L$, with $L \rightarrow \infty$.
When the $N=3$ relays operate in FD, they all have the same single capacity given by $ \mathsf{C}^{{\rm{FD}}}_{\mcal{N}_{\{i\}}} = \ell, \forall i \in [1:3]$.
This means that, by selecting any of the relays (i.e., at random), we get the same performance guarantee. 
Differently, when the $N=3$ relays operate in HD, the third relay is better giving 
$\mathsf{C}_{\mcal{N}_{\{3\}}} =\ell$.
These two simple examples suggest that, when the relays operate in HD, choosing the best subnetwork based on the FD capacities might not be a smart choice.
For instance, in the second example if we select either the first or the second relay (which in FD are optimal) we would incur a loss of $50 \%$ in the approximate capacity (which is also the maximum loss value) compared to selecting the third relay.

\smallskip
\noindent{\bf{3) Worst-case networks in HD and FD are not necessarily the same.}}
Consider the network example in~\eqref{eq:WorstNetwEx} and suppose we want to select $N-1$ relays.
We already showed (see Section~\ref{sec:FundGar}) that, by selecting any $(N-1)$-relay subnetwork $\mathcal{N}_\mathcal{K}$ with $|\mathcal{K}|=N-1$, we get
$\mathsf{C}_{\mathcal{N}_{\mathcal{K}}} = \frac{N-1}{N}{\mathsf{C}}_{\mathcal{N}_{[1:N]}}$,  
i.e., the network in~\eqref{eq:WorstNetwEx}, when operated in HD, represents a worst-case scenario.
Now, suppose that we operate the network in~\eqref{eq:WorstNetwEx} in FD. 
Then, it is not difficult to see that there always exists an $(N-1)$-relay subnetwork $\mathcal{N}_{\mathcal{K}}$ with $|\mathcal{K}|=N-1$, that guarantees 
$\mathsf{C}^{{\rm{FD}}}_{\mathcal{N}_{\mathcal{K}}} = {\mathsf{C}}^{{\rm{FD}}}_{\mathcal{N}_{[1:N]}}$,
which is greater than the worst-case ratio of~$\frac{N-1}{N}$ proved in~\cite[Theorem 1]{NazarogluIT2014}.
This suggests that tight network examples for HD with general values of $k$ and $N$ might not be the same as those in FD; this adds an extra degree of complication in the study of the network simplification problem in HD since the approximate capacity in HD (because of the required optimization over the $2^N$ listen/transmit configuration states) cannot be computed directly as in the FD counterpart.

\smallskip

In this paper, we investigated the network simplification problem in an $N$-relay Gaussian HD diamond network.
We proved that there always exists a subnetwork of $k=N-1$ relays that achieves at least a fraction $\frac{N-1}{N}$ of the approximate capacity of the full network.
This result was derived by showing that any optimal schedule of the full network can be used by at least one of the $N$ subnetworks of $k=N-1$ relays to achieve the worst performance guarantee.
Moreover, we provided an example of a class of Gaussian HD diamond networks for which this fraction is tight.
Then, by leveraging the results obtained for $k=N-1$, we derived lower bounds on the fraction guarantee for general $k \in [1:N]$, which are tighter than currently available bounds when $k > \frac{N}{2}$.
Finally, we showed that, when we select $k=1$ or $k=2$ relays, the fraction guarantee decreases as $N$ increases; this is a surprising difference between the network simplication problem in HD and FD.
These results were obtained by leveraging properties of submodular functions and diamond networks that were derived here and that might be of independent interest for other applications.

\begin{appendices}
\section{Proof of Lemma~\ref{lem:part}}
\label{app:lemmaPart}
In order to prove the result in Lemma~\ref{lem:part}, we make use of the following lemma, valid for Gaussian FD diamond networks.
\begin{lem}
\label{lemma:FDpart}
For any Gaussian FD diamond network $\mathcal{N}_{[1:N]}$ and  $\mcal{K} \subseteq [1:N]$, we have that
\begin{align}
{\mathsf{C}}^{\rm{FD}}_{\mathcal{N}_{[1:N]}}
 \leq{\mathsf{C}}^{\rm{FD}}_{\mathcal{N}_{\mathcal{K}}} +{\mathsf{C}}^{\rm{FD}}_{\mathcal{N}_{[1:N]\backslash \mathcal{K}}},
\end{align}
where
\begin{align*}
&{\mathsf{C}}^{\rm{FD}}_{\mathcal{N}_{[1:N]}}
= \min_{\mathcal{A}_{\rm{F}} \subseteq [1:N]} \left \{
\max_{i \in \mathcal{A}_{\rm{F}}} \ell_i + \max_{i \in \mathcal{N}_{[1:N]} \backslash \mathcal{A}_{\rm{F}}} r_i
\right \} 
\\
& {\mathsf{C}}^{\rm{FD}}_{\mcal{N}_{\mathcal{M}_j}} 
= \min_{\mathcal{A}_j \subseteq \mathcal{M}_j} \left \{ \max_{i \in \mathcal{A}_j} \ell_i +\max_{i \in \mathcal{M}_j \backslash \mathcal{A}_j} r_i \right \}, 
\qquad  \forall j \in [1:2],
\end{align*}
with $\mathcal{M}_1=\mathcal{K}$ and $\mathcal{M}_2=[1:N]\backslash \mathcal{K}$.
\end{lem}

\begin{IEEEproof}
We have
\begin{align}
&\max_{i \in \mathcal{A}_1} \ell_i+\max_{i \in \mathcal{A}_2} \ell_i + \max_{i \in \mathcal{M}_1 \backslash \mathcal{A}_1} r_i + \max_{i \in \mathcal{M}_2 \backslash \mathcal{A}_2} r_i  \nonumber
\\  \geq &  \max_{i \in \mathcal{A}_1 \cup \mathcal{A}_2} \ell_i+ \max_{i \in \left(\mathcal{M}_1 \backslash \mathcal{A}_1\right) \cup \left(\mathcal{M}_2 \backslash \mathcal{A}_2\right)} r_i \nonumber
\\ \stackrel{{\rm{(a)}}}{=} &
\max_{i \in \mathcal{A}_1 \cup \mathcal{A}_2} \ell_i+ \max_{i \in \left(\mathcal{M}_1 \cup \mathcal{M}_2 \right) \backslash{ \left(\mathcal{A}_1 \cup \mathcal{A}_2 \right)}} r_i \nonumber
 \\=&  \max_{i \in \mathcal{A}_{\rm{F}}} \ell_i + \max_{i \in{ [1:N]} \backslash \mathcal{A}_{\rm{F}}} r_i \nonumber
\\ \geq & \min_{\mathcal{A}_{\rm{F}}\subseteq { [1:N]}} \left \{\max_{i \in \mathcal{A}_{\rm{F}}} \ell_i + \max_{i \in {[1:N]} \backslash \mathcal{A}_{\rm{F}}} r_i \right \}
= \mathsf{C}_{\mathcal{N}_{[1:N]}}^{\rm{FD}}.
\label{eq:boundsFDpartition}
\end{align}
The equality in $\rm{(a)}$ appeals to the following property (recall that $\mathcal{M}_1$ and $\mathcal{M}_2$ are disjoint and $\mathcal{A}_i \subseteq \mathcal{M}_i, i \in [1:2]$)
\begin{align*}
 \left(\mathcal{M}_1 \backslash \mathcal{A}_1\right) \cup \left(\mathcal{M}_2 \backslash \mathcal{A}_2\right) & \stackrel{{\rm{(b)}}}{=}  \left (\mathcal{M}_1 \backslash \left(\mathcal{A}_1 \cup \mathcal{A}_2 \right) \right) \cup \left (\mathcal{M}_2 \backslash \left(\mathcal{A}_1 \cup \mathcal{A}_2 \right) \right)
 \\& \stackrel{{\rm{(c)}}}{=} \left( \mathcal{M}_1 \cup \mathcal{M}_2 \right) \backslash \left( \mathcal{A}_1 \cup \mathcal{A}_2 \right),
\end{align*}
where the equality in $\rm{(b)}$ follows since $\mathcal{M}_1 \cap \mathcal{A}_2 = \emptyset$ and $\mathcal{M}_2 \cap \mathcal{A}_1 = \emptyset$ and
the equality in $\rm{(c)}$ follows since $\left( \mathcal{B} \backslash \mathcal{A} \right) \cup \left( \mathcal{C} \backslash \mathcal{A} \right) = \left( \mathcal{B} \cup \mathcal{C}\right) \backslash \mathcal{A}$.
The result in \eqref{eq:boundsFDpartition} is valid $\forall \mathcal{A}_1 \subseteq \mathcal{M}_1$ and $\forall \mathcal{A}_2 \subseteq \mathcal{M}_2$, hence also for the minimum cuts of the networks {$\mathcal{N}_{\mathcal{M}_1}$ and $\mathcal{N}_{\mathcal{M}_2}$, i.e., }
\begin{align*}
{\mathsf{C}}^{\rm{FD}}_{\mathcal{N}_{\mathcal{M}_1}} +{\mathsf{C}}^{\rm{FD}}_{\mathcal{N}_{\mathcal{M}_2}} = 
{\mathsf{C}}^{\rm{FD}}_{\mathcal{N}_{\mathcal{K}}} +{\mathsf{C}}^{\rm{FD}}_{\mathcal{N}_{[1:N]\backslash \mathcal{K}}}
\geq
\mathsf{C}_{ \mathcal{N}_{[1:N]}}^{\rm{FD}}.
\end{align*}
\end{IEEEproof}
\smallskip
We now show how the result in Lemma~\ref{lemma:FDpart}, valid for Gaussian FD diamond networks, extends to the HD case.
For a given schedule $\lambda$ of the full network $\mathcal{N}_{[1:N]}$, we have from~\eqref{eq:subnetwork_using_full_schedule} that
\begin{align*}
{{\mathsf{R}}^{\lambda}_{ \mathcal{N}_{[1:N]}}}
 & = \min_{\mathcal{A}_{\rm{F}} \subseteq {[1:N]}} \sum_{s\in [0:1]^N} \lambda_s \left( \max_{i \in\mathcal{A}_{\rm{F}} }  \ell_{i,s}^{\prime} + \max_{i \in {[1:N]} \backslash \mathcal{A}_{\rm{F}} }  r_{i,s}^{\prime}\right),
\end{align*}
where $\ell_{i,s}^{\prime}$ and $r_{i,s}^{\prime}$ are defined in~\eqref{eq:NewChannels}.
From the result in~\eqref{eq:boundsFDpartition}, $\forall \mathcal{A}_1 \subseteq \mathcal{M}_1$ and $\forall \mathcal{A}_2 \subseteq \mathcal{M}_2$, with {$\mathcal{M}_1=\mathcal{K}$ and $\mathcal{M}_2=[1:N]\backslash \mathcal{K}$,}
we have that
\begin{align*}
&\sum_{s\in [0:1]^N} \lambda_s\left [ 
   \max_{i \in\mathcal{A}_1 }  \ell_{i,s}^{\prime}
 + \max_{i \in\mathcal{A}_2 }  \ell_{i,s}^{\prime}
 + \max_{i \in \mathcal{M}_1 \backslash \mathcal{A}_1 }  r_{i,s}^{\prime}
 + \max_{i \in \mathcal{M}_2 \backslash \mathcal{A}_2 }  r_{i,s}^{\prime}
\right ]
\\ \geq & \sum_{s\in [0:1]^N} \lambda_s \left( 
  \max_{i \in\mathcal{A}_{\rm{F}} }  \ell_{i,s}^{\prime}
+ \max_{i \in {[1:N]} \backslash \mathcal{A}_{\rm{F}} }  r_{i,s}^{\prime} \right) \geq 
{{\mathsf{R}}^{\lambda}_{ \mathcal{N}_{[1:N]}}},
\end{align*}
where $\mathcal{A}_{\rm{F}} = \mathcal{A}_1 \cup \mathcal{A}_2$. 
This implies 
\begin{align*}
{{\mathsf{R}}^{\lambda}_{\mathcal{N}_{[1:N]}} \leq {\mathsf{R}}^{\lambda}_{\mathcal{N}_{\mathcal{M}_{1}}} +{\mathsf{R}}^{\lambda}_{\mathcal{N}_{\mathcal{M}_{2}}}
=
{\mathsf{R}}^{\lambda}_{\mathcal{N}_{\mathcal{K}}} +{\mathsf{R}}^{\lambda}_{\mathcal{N}_{[1:N]\backslash \mathcal{K}}}}.
\end{align*}
This concludes the proof of Lemma~\ref{lem:part}.

\section{Proof of Lemma~\ref{lemma:subProperty} } 
\label{appendix:proof_of_subproperty}
Let $f$ be a submodular set function defined on $\Omega$ (see Definition~\ref{def:subFunc}). 
We want to prove that for any collection of $n$ sets $\mathcal{A}_i \subseteq \Omega$,
\begin{align*}
\sum_{i=1}^n f \left( \mathcal{A}_i \right) \geq
\sum_{j=1}^n f \left( \mathcal{E}^{(n)}_j \right),
\end{align*}
where $\mathcal{E}_j^{(n)}$ is the set of elements that appear in at least $j$ sets $\mathcal{A}_i, i \in [1:n]$.
The proof is by induction.
For the base case (i.e., $n=1$) we clearly have that $f(\mathcal{A}_1) = f \left (\mathcal{E}^{(1)}_1 \right)$.
For the proof of the induction step, we prove and use the following property of submodular functions.
\begin{prope}
\label{prope:subfun}
Let $f$ be a submodular function. Then, $\forall n>0$ and $0 \leq k<n$,
\begin{align}
\label{eq:subProp}
&f \left( \bigcup_{\substack{\mathcal{I} \subseteq [1:n] \\ |\mathcal{I}|=k}}  \left( \mathcal{A}_{n+1}\bigcap_{i \in \mathcal{I}} \mathcal{A}_i \right)\right)
+ f \left( \bigcup_{\substack{\mathcal{I} \subseteq [1:n]\\ |\mathcal{I}|=k+1}} \left(\bigcap_{i \in \mathcal{I}} \mathcal{A}_i \right)\right) \nonumber
\\ &\geq f\left(\bigcup_{\substack{\mathcal{I} \subseteq [1:n+1] \\ |\mathcal{I}|=k+1}} \left(\bigcap_{i \in \mathcal{I}} \mathcal{A}_i \right)\right)
+  f \left(  \bigcup_{\substack{\mathcal{I} \subseteq [1:n]\\ |\mathcal{I}|=k+1}} \left(\mathcal{A}_{n+1} \bigcap_{i \in \mathcal{I}} \mathcal{A}_i \right) \right).
\end{align}
\end{prope}
We now use Property~\ref{prope:subfun}, whose proof can be found at the end of this appendix, to prove the induction step.
Assume that for some $n>0$, we have that
\begin{align}
\label{eq:StartInd}
\sum_{i=1}^n f \left( \mathcal{A}_i \right) \geq
\sum_{j=1}^n f \left( \mathcal{E}_j^{(n)} \right).
\end{align}
Our goal is to prove that
\begin{align*}
\sum_{i=1}^{n+1} f \left( \mathcal{A}_i \right) \geq
\sum_{j=1}^{n+1} f \left( \mathcal{E}_j^{(n+1)} \right).
\end{align*}
From~\eqref{eq:StartInd}, by adding the positive quantity $f\left(\mathcal{A}_{n+1} \right)$ to both sides of the inequality, we have that
\begin{align*}
\sum_{i=1}^n f \left( \mathcal{A}_i \right) + f\left(\mathcal{A}_{n+1} \right)\geq
\sum_{j=1}^n f \left( \mathcal{E}_j^{(n)} \right)+ f\left(\mathcal{A}_{n+1} \right),
\end{align*}
which can be equivalently rewritten as
\begin{align*}
\sum_{i=1}^n f \left( \mathcal{A}_i \right) + f\left(\mathcal{A}_{n+1} \right) &\geq
f\left(\mathcal{A}_{n+1} \right) + \sum_{k=0}^{n-1} 
\underbrace{ f \left( \bigcup_{\substack{\mathcal{I}\subseteq [1:n]\\ |\mathcal{I}|=k+1}} \left( \bigcap_{i \in \mathcal{I}} \mathcal{A}_i\right) \right)}_{f \left( \mathcal{E}_{k+1}^{(n)}\right)}
\\&=f\left(\mathcal{A}_{n+1} \right) + f \left( \bigcup_{\substack{\mathcal{I}\subseteq [1:n]\\ |\mathcal{I}|=1}} \left( \bigcap_{i \in \mathcal{I}} \mathcal{A}_i\right) \right) + \sum_{k=1}^{n-1} f \left( \bigcup_{\substack{\mathcal{I}\subseteq [1:n]\\ |\mathcal{I}|=k+1}} \left( \bigcap_{i \in \mathcal{I}} \mathcal{A}_i\right) \right).
\end{align*}
The final step in the proof follows by inductively applying Property~\ref{prope:subfun} on the underlined terms with the appropriate $k$ as shown in what follows,
\begin{align*} 
& \underline{f\left(\mathcal{A}_{n+1} \right) 
+ f \left( \bigcup_{\substack{\mathcal{I}\subseteq [1:n]\\ |\mathcal{I}|=1}} \left( \bigcap_{i \in \mathcal{I}} \mathcal{A}_i\right) \right)}
+ \sum_{k=1}^{n-1} f \left( \bigcup_{\substack{\mathcal{I}\subseteq [1:n]\\ |\mathcal{I}|=k+1}} \left( \bigcap_{i \in \mathcal{I}} \mathcal{A}_i\right) \right) 
\\ &\stackrel{(k = 0)}{\geq}
f \left( \bigcup_{\substack{\mathcal{I}\subseteq [1:n+1]\\ |\mathcal{I}|=1}} \left( \bigcap_{i \in \mathcal{I}} \mathcal{A}_i\right) \right)
+ \underline{f \left( \bigcup_{\substack{\mathcal{I}\subseteq [1:n]\\ |\mathcal{I}|=1}} \left( \mathcal{A}_{n+1} \bigcap_{i \in \mathcal{I}} \mathcal{A}_i\right) \right)
+ f \left( \bigcup_{\substack{\mathcal{I}\subseteq [1:n]\\ |\mathcal{I}|=2}} \left( \bigcap_{i \in \mathcal{I}} \mathcal{A}_i\right) \right)} 
\\& \qquad + \sum_{k=2}^{n-1} f \left( \bigcup_{\substack{\mathcal{I}\subseteq [1:n]\\ |\mathcal{I}|=k+1}} \left( \bigcap_{i \in \mathcal{I}} \mathcal{A}_i\right) \right)
\\ &\stackrel{(k = 1)}{\geq}
\sum_{\ell=1}^2 f \left( \bigcup_{\substack{\mathcal{I}\subseteq [1:n+1]\\ |\mathcal{I}|=\ell}} \left( \bigcap_{i \in \mathcal{I}} \mathcal{A}_i\right) \right)
+ \underline{f \left( \bigcup_{\substack{\mathcal{I}\subseteq [1:n]\\ |\mathcal{I}|=2}} \left( \mathcal{A}_{n+1} \bigcap_{i \in \mathcal{I}} \mathcal{A}_i\right) \right)
+ f \left( \bigcup_{\substack{\mathcal{I}\subseteq [1:n]\\ |\mathcal{I}|=3}} \left( \bigcap_{i \in \mathcal{I}} \mathcal{A}_i\right) \right)}
\\& \qquad + \sum_{k=3}^{n-1} f \left( \bigcup_{\substack{\mathcal{I}\subseteq [1:n]\\ |\mathcal{I}|=k+1}} \left( \bigcap_{i \in \mathcal{I}} \mathcal{A}_i\right) \right) 
\\& \qquad \qquad \qquad\qquad \qquad \qquad \qquad\qquad   \vdots 
\\ &\stackrel{(k=n-1)}{\geq} 
 \sum_{\ell=1}^n f  \left(  \bigcup_{\substack{\mathcal{I}\subseteq [1:n+1]\\ |\mathcal{I}|=\ell}} \left( \bigcap_{i \in \mathcal{I}} \mathcal{A}_i\right) \right)
+  f \left(  \bigcup_{\substack{\mathcal{I}\subseteq [1:n]\\ |\mathcal{I}|=n}} \left( \mathcal{A}_{n+1} \bigcap_{i \in \mathcal{I}} \mathcal{A}_i\right) \right)
\\& =
\sum_{\ell=1}^{n+1} f  \left(  \bigcup_{\substack{\mathcal{I}\subseteq [1:n+1]\\ |\mathcal{I}|=\ell}} \left( \bigcap_{i \in \mathcal{I}} \mathcal{A}_i\right) \right) =  \sum_{j=1}^{n+1} f  \left( \mathcal{E}^{(n+1)}_j \right).
\end{align*}
This concludes the proof of Lemma~\ref{lemma:subProperty}.

\subsection{Proof of Property~\ref{prope:subfun}}
By using properties of submodular functions and set operations we have
\begin{align*}
& f \left( \bigcup_{\substack{\mathcal{I} \subseteq [1:n] \\ |\mathcal{I}|=k}} \left( \mathcal{A}_{n+1} \bigcap_{i \in \mathcal{I}} \mathcal{A}_i \right)\right)
+ f \left( \bigcup_{\substack{\mathcal{I} \subseteq [1:n]\\ |\mathcal{I}|=k+1}} \left(\bigcap_{i \in \mathcal{I}} \mathcal{A}_i \right)\right)
\\\stackrel{{\rm{(a)}}}{\geq} & f \left(\left [ \bigcup_{\substack{\mathcal{I} \subseteq [1:n] \\ |\mathcal{I}|=k}} \left( \mathcal{A}_{n+1} \bigcap_{i \in \mathcal{I}} \mathcal{A}_i \right) \right ] \bigcup \left [ \bigcup_{\substack{\mathcal{I} \subseteq [1:n]\\ |\mathcal{I}|=k+1}} \left(\bigcap_{i \in \mathcal{I}} \mathcal{A}_i \right)\right ] \right)
\\ & + f \left(  \left [ \bigcup_{\substack{\mathcal{I} \subseteq [1:n] \\ |\mathcal{I}|=k}} \left( \mathcal{A}_{n+1} \bigcap_{i \in \mathcal{I}} \mathcal{A}_i \right)\! \right] \bigcap  \left [ \bigcup_{\substack{\mathcal{I} \subseteq [1:n]\\ |\mathcal{I}|=k+1}} \left(\bigcap_{i \in \mathcal{I}} \mathcal{A}_i \right)\right ]  \right)
\\ \stackrel{{\rm{(b)}}}{=}& f \left( \bigcup_{\substack{\mathcal{I} \subseteq [1:n+1] \\ |\mathcal{I}|=k+1}} \left(\bigcap_{i \in \mathcal{I}} \mathcal{A}_i \right)\right) + f \left(  \left [ \bigcup_{\substack{\mathcal{I} \subseteq [1:n] \\ |\mathcal{I}|=k}} \left( \mathcal{A}_{n+1} \bigcap_{i \in \mathcal{I}} \mathcal{A}_i \right)\! \right] \bigcap  \left [ \bigcup_{\substack{\mathcal{I} \subseteq [1:n]\\ |\mathcal{I}|=k+1}} \left(\bigcap_{i \in \mathcal{I}} \mathcal{A}_i \right)\right ]  \right)
\\ \stackrel{{\rm{(c)}}}{=}& \underbrace{f \left( \bigcup_{\substack{\mathcal{I} \subseteq [1:n+1] \\ |\mathcal{I}|=k+1}} \left(\bigcap_{i \in \mathcal{I}} \mathcal{A}_i \right)\right)}_{T_1}+ 
f\left(\underbrace{\mathcal{A}_{n+1}\bigcap \left[ \bigcup_{\substack{\mathcal{J} \subseteq [1:n] \\ |\mathcal{J}|=k}} \bigcap_{j \in \mathcal{J}} \mathcal{A}_j \right] \bigcap \left[ \bigcup_{\substack{\mathcal{I}\subseteq [1:n] \\ |\mathcal{I}|=k+1}}  \bigcap_{i \in \mathcal{I}} \mathcal{A}_i \right ]}_{\mathcal{S}} \right) ,
\end{align*}
where: 
(i) the inequality in ${\rm{(a)}}$ follows from the definition of submodular function (see Definition~\ref{def:subFunc});
(ii) the equality in $\rm{(b)}$ follows by combining the union in the first term of the inequality in ${\rm{(a)}}$;
(iii) the equality in $\rm{(c)}$ follows from the distributive property of intersection over unions.
Note that $T_1$ is already the first term we need in the inequality.
To arrive at the second term, we shall prove that
\begin{align} 
\label{eq:set_S}
\mathcal{S} = \bigcup_{\substack{\mathcal{I} \subseteq [1:n]\\ |\mathcal{I}|=k+1}} \left(\mathcal{A}_{n+1} \bigcap_{i \in \mathcal{I}} \mathcal{A}_i \right)  .
\end{align}
Towards this end, notice that the distributive property of intersection over unions gives
\begin{align} 
\label{eq:sets_induction}
&\mathcal{A}_{n+1}\bigcap\left[\bigcup_{\substack{\mathcal{J} \subseteq [1:n] \\ |\mathcal{J}|=k}} \bigcap_{j \in \mathcal{J}} \mathcal{A}_j\right]\bigcap\left[\bigcup_{\substack{\mathcal{I}\subseteq [1:n] \\ |\mathcal{I}|=k+1}}  \bigcap_{i \in \mathcal{I}} \mathcal{A}_i \right] \nonumber
\\
&=\mathcal{A}_{n+1}\bigcap\bigcup_{\substack{\mathcal{I}\subseteq [1:n] \\ |\mathcal{I}|=k+1}} \left[ \left(\bigcap_{i \in \mathcal{I}} \mathcal{A}_i \right) \bigcap\left( \bigcup_{\substack{\mathcal{J} \subseteq [1:n] \\ |\mathcal{J}|=k}} \bigcap_{j \in \mathcal{J}} \mathcal{A}_j \right) \right].
\end{align}
Now note that $\forall \mathcal{I} \subseteq [1:n]$ with $|\mathcal{I}|=k+1$, $\exists\;\mathcal{J}_{\mathcal{I}} \subset \mathcal{I}$ with $|\mathcal{J}_{\mathcal{I}}|=k$. 
This observation implies that, for each $\mathcal{I}$, we have
\begin{align*}
&\left( \bigcap_{i \in \mathcal{J}_\mathcal{I} } \mathcal{A}_i \right) \bigcap \left( \bigcup_{\substack{\mathcal{J} \subseteq [1:n] \\ |\mathcal{J}|=k}} \bigcap_{j \in \mathcal{J}} \mathcal{A}_j \right)
=\left( \bigcap_{i \in \mathcal{J}_\mathcal{I}} \mathcal{A}_i \right) \bigcap \left( \left(  \bigcap_{i \in \mathcal{J}_\mathcal{I}} \mathcal{A}_i \right) \bigcup \left(\bigcup_{\substack{\mathcal{L} \subseteq [1:n]\\ \mathcal{L} \neq \mathcal{J}_\mathcal{I}\\ |\mathcal{L}|=k}} \bigcap_{\ell \in \mathcal{L}} \mathcal{A}_\ell \right)\right)
\stackrel{{\rm{(c)}}}{=} \bigcap_{i \in \mathcal{J}_\mathcal{I}} \mathcal{A}_i,
\end{align*}
where the equality in ${\rm{(c)}}$ follows since $\mathcal{U} \cap (\mathcal{U} \cup \mathcal{V}) = \mathcal{U}$.
As a consequence, for each $\mathcal{I}$, we have
\begin{align} 
\label{eq:set_intersection_union_prop}
\left( \bigcap_{i \in \mathcal{I}} \mathcal{A}_i \right) \bigcap \left( \bigcup_{\substack{\mathcal{J} \subseteq [1:n] \\ |\mathcal{J}|=k}} \bigcap_{j \in \mathcal{J}} \mathcal{A}_j \right) \nonumber
&= \left( \bigcap_{i \in \mathcal{I}\backslash \mathcal{J}_\mathcal{I} } \mathcal{A}_i \right)\bigcap \left( \bigcap_{i \in \mathcal{J}_\mathcal{I} } \mathcal{A}_i \right) \bigcap \left( \bigcup_{\substack{\mathcal{J} \subseteq [1:n] \\ |\mathcal{J}|=k}} \bigcap_{j \in \mathcal{J}} \mathcal{A}_j \right) \nonumber
\\
& = \left( \bigcap_{i \in \mathcal{I}\backslash \mathcal{J}_\mathcal{I} } \mathcal{A}_i \right)\bigcap \left( \bigcap_{i \in \mathcal{J}_\mathcal{I} } \mathcal{A}_i \right)  = \left( \bigcap_{i \in \mathcal{I}} \mathcal{A}_i \right).
\end{align}
Finally, by applying~\eqref{eq:set_intersection_union_prop} for each $\mathcal{I}$ in~\eqref{eq:sets_induction}, we get
\begin{align*}
\mathcal{S} &= \mathcal{A}_{n+1}\bigcap\left[ \bigcup_{\substack{\mathcal{J} \subseteq [1:n] \\ |\mathcal{J}|=k}} \bigcap_{j \in \mathcal{J}} \mathcal{A}_j \right] \bigcap \left[ \bigcup_{\substack{\mathcal{I}\subseteq [1:n] \\ |\mathcal{I}|=k+1}}  \bigcap_{i \in \mathcal{I}} \mathcal{A}_i \right ] \\
& =\mathcal{A}_{n+1} \bigcap \bigcup_{\substack{\mathcal{I}\subseteq [1:n] \\ |\mathcal{I}|=k+1}} \left[ \left( \bigcap_{i \in \mathcal{I}} \mathcal{A}_i \right) \bigcap \left( \bigcup_{\substack{\mathcal{J} \subseteq [1:n] \\ |\mathcal{J}|=k}} \bigcap_{j \in \mathcal{J}} \mathcal{A}_j \right) \right]
\\
&= \mathcal{A}_{n+1} \bigcap\left(\bigcup_{\substack{\mathcal{I}\subseteq [1:n] \\ |\mathcal{I}|=k+1}} \bigcap_{i \in \mathcal{I}} \mathcal{A}_i \right)
    =\bigcup_{\substack{\mathcal{I} \subseteq [1:n]\\ |\mathcal{I}|=k+1}} \left(\mathcal{A}_{n+1} \bigcap_{i \in \mathcal{I}} \mathcal{A}_i \right),
\end{align*}
where the last equality follows by using the distributive property of intersection over unions.
This proves~\eqref{eq:set_S} hence concluding the proof of Property~\ref{prope:subfun}.

\section{Proof of Lemma~\ref{Lemma:FD3From4}}
\label{app:LemmaDiamond}
From the statement of Lemma~\ref{Lemma:FD3From4}, recall that $\mathcal{A}_i \subseteq [1:N] \backslash \{i\}$.
Throughout the proof, we let 
$\mathcal{B}_i = ([1:N]\backslash \{i\}) \backslash \mathcal{A}_i, \ \forall i \in [1:N]$,
$f \left( \mathcal{A} \right) = \max_{i \in \mathcal{A}} \ell_i$ and $g \left( \mathcal{A} \right) = \max_{i \in \mathcal{A}} r_i$, with $\mathcal{A} \subseteq [1:N]$.
It is not difficult to see that $f$ and $g$ are submodular functions.
As a result, we have
\begin{align}
&\sum_{j=1}^N \left(\max_{i \in \mathcal{A}_j} \ell_i  
+ \max_{i \in ([1:N]\backslash \{j\})  \backslash \mathcal{A}_j} r_i \right) \nonumber
\\ = & \sum_{j=1}^N \left[ f \left( \mathcal{A}_j \right) + g \left( \mathcal{B}_j \right) \right] \nonumber
\\ \stackrel{{\rm{(a)}}}{\geq}&
\sum_{j=1}^N \left [f \left( \mathcal{E}_j^{(N)} \right)  + g \left( \mathcal{F}_j^{(N)} \right)  \right ] \nonumber
 \\ \stackrel{{\rm{(b)}}}{=} &
\sum_{j=1}^{N-1} \left [f \left( \mathcal{E}_j^{(N)} \right)  + g \left( \mathcal{F}_j^{(N)} \right)  \right ] \nonumber
\\ \stackrel{{\rm{(c)}}}{=}&
\sum_{j=1}^{N-1} \left [f \left( \mathcal{E}_j^{(N)} \right)  + g \left( \mathcal{F}_{N-j}^{(N)} \right)  \right ],
\label{eq:almostDone}
\end{align}
where: 
(i) the inequality in $\rm{(a)}$ follows from Lemma~\ref{lemma:subProperty} with $\mathcal{E}_j^{(N)}$ (respectively, $\mathcal{F}_j^{(N)}$) being the set of elements that appear in at least $j$ sets $\mathcal{A}_i, \ i \in [1:N]$ (respectively, $\mathcal{B}_i$);
(ii) the equality in $\rm{(b)}$ follows because $\mathcal{E}^{(N)}_N = \mathcal{F}^{(N)}_N=\emptyset$ since $\bigcap_{i=1}^N ([1:N] \backslash \{i\}) = \emptyset$; 
(iii) the equality in ${\rm (c)}$ follows by simply reordering the sum.

Note that $\forall i \in [1:N]$, the element $i \in [1:N]\backslash \{j\}$, with $j\neq i$, and $\mathcal{A}_j$ and $\mathcal{B}_j$ are by definition disjoint $\forall j \in [1:N]$.
Thus, the element $i$ belongs to exactly $(N-1)$ sets $\mathcal{A}_j$, $\mathcal{B}_j$.
We now claim that $[1:N] \backslash \mathcal{E}_j^{(N)} = \mathcal{F}_{N-j}^{(N)}, \ j \in [1:N-1]$.
Consider an element $x \in [1:N]$; then:
\begin{enumerate}
\item Let $x \in \mathcal{E}_j^{(N)}$, i.e., $x$ appears in at least $j$ sets $\mathcal{A}_i$. Since $x$ appears exactly $(N-1)$ times in $\mathcal{A}_i$ and $\mathcal{B}_i$, this means that $x$ appears in at most $(N-1)-j$ sets $\mathcal{B}_i$, i.e., $x \notin \mathcal{F}_{N-j}^{(N)}$. 
In other words, $x \in [1:N] \backslash \mathcal{F}_{N-j}^{(N)}$. 
Since this is true $\forall x \in \mathcal{E}_j^{(N)}$, it implies that $\mathcal{E}_j^{(N)} \subseteq [1:N] \backslash \mathcal{F}^{(N)}_{N-j}$ and as a result $[1:N] \backslash \mathcal{E}_j^{(N)} \supseteq \mathcal{F}_{N-j}^{(N)}$.
\item Let $x \notin \mathcal{E}_j^{(N)}$, i.e., $x$ appears in at most $(j-1)$ sets $\mathcal{A}_i$; since $x$ in total appears exactly $(N-1)$ times in $\mathcal{A}_i$ and $\mathcal{B}_i$, this means that $x$ appears in at least $(N-1)-(j-1)$ sets $\mathcal{B}_i$, i.e., $x \in \mathcal{F}_{N-j}^{(N)}$. 
Since this is true $\forall x \in [1:N]\backslash\mathcal{E}_j^{(N)}$, it implies that $[1:N] \backslash \mathcal{E}_j^{(N)} \subseteq \mathcal{F}_{N-j}^{(N)}$.
\end{enumerate}
The points in 1) and 2) imply that $[1:N] \backslash \mathcal{E}_j^{(N)} = \mathcal{F}_{N-j}^{(N)},\ \forall j \in [1:N-1]$. 
Applying this equality into~\eqref{eq:almostDone}, we obtain
\begin{align*}
&\sum_{j=1}^N \left(\max_{i \in \mathcal{A}_j} \ell_i 
+ \max_{i \in ([1:N] \backslash \{j\}) \backslash \mathcal{A}_j} r_i \right) 
\geq  
\sum_{j=1}^{N-1} \left [f \left( \mathcal{E}_j^{(N)} \right)  + g \left( \mathcal{F}_{N-j}^{(N)} \right)  \right ]
\\ = & \sum_{j=1}^{N-1} \left [f \left( \mathcal{E}_j^{(N)} \right)  + g \left( [1:N] \backslash \mathcal{E}_j^{(N)} \right)  \right ]
  =   \sum_{j=1}^{N-1} 
\left(\max_{i \in \mathcal{A}_{{\rm{F}}j}} \ell_i 
+ \max_{i \in [1:N] \backslash \mathcal{A}_{{\rm{F}}j}} r_i \right),
\end{align*}
where we let $\mathcal{A}_{{\rm{F}}j}= \mathcal{E}_j^{(N)}$.
Since throughout the proof we made no assumptions on the values of $(\ell_i,\ r_i)$, then the sets $\mcal{A}_{{\rm F}j}$ do not depend on the values of $(\ell_i,r_i)$.
This concludes the proof of Lemma~\ref{Lemma:FD3From4}.

\section{Proof of Lemma~\ref{Lemma:HD3From4}}
\label{app:proof_mainresults}
Let $\lambda$ be a schedule (non necessarily optimal) of the full network $\mathcal{N}_{[1:N]}$ with $N$ relays. 
Denote by $\mcal{A}_j^\star$ the minimum cut of the network $\bar{\mcal{N}}_j$ when operated with the `natural' schedule constructed from $\lambda$.
Then, by following the same steps as in the example in Section~\ref{sec:system_model}, from~\eqref{eq:subnetwork_using_full_schedule} we obtain
\begin{align*}
    \sum_{i=1}^N {\mathsf{R}}^{\lambda}_{\bar{\mathcal{N}}_i} =
\sum_{s\in [0:1]^N} \lambda_s
\left [
\sum_{j=1}^N 
\left(
\max_{i \in \mathcal{A}^{\star}_j} \ell_{i,s}^{\prime} +\max_{i \in ([1:N]\backslash \{j\})  \backslash  \mathcal{A}^{\star}_j} r_{i,s}^{\prime} 
\right)
\right],
\end{align*}
where $\ell_{i,s}^{\prime}$ and $r_{i,s}^{\prime} $ are defined in~\eqref{eq:NewChannels}.
From the result in Lemma~\ref{Lemma:FD3From4} we know that $\exists \left \{\mathcal{A}_{{\rm{F}}j} \right \}, \ j \in [1:N-1]$, such that for each $s \in [0:1]^N$:
\begin{align*}
&\sum_{j=1}^N
\left(
\max_{i \in \mathcal{A}^{\star}_j} \ell_{i,s}^{\prime} +\max_{i \in ([1:N]\backslash \{j\})  \backslash  \mathcal{A}^{\star}_j} r_{i,s}^{\prime} 
\right)\geq 
\sum_{j=1}^{N-1}
\left(\max_{i \in \mathcal{A}_{{\rm{F}}j}} \ell_{i,s}^{\prime}  
+ \max_{i \in [1:N] \backslash \mathcal{A}_{{\rm{F}}j}} r_{i,s}^{\prime} \right),
\end{align*}
where $\mathcal{A}_{{\rm{F}}j} \subseteq [1:N], \ \forall j \in [1:N-1]$. 
Additionally, from Lemma~\ref{Lemma:FD3From4} we have that $\mcal{A}_{{\rm{F}}j}$ is independent of $(\ell_{i,s}^\prime, r_{i,s}^\prime)$ and is therefore independent of $(\ell_i,r_i)$ and of the relaying state $s$. 
Hence
\begin{align*}
\sum_{i=1}^N {\mathsf{R}}^{\lambda}_{\bar{\mathcal{N}}_i}  &=
\sum_{s\in [0:1]^N} \lambda_s
\left [
\sum_{j=1}^N 
\left(
\max_{i \in \mathcal{A}^{\star}_j} \ell_{i,s}^{\prime} +\max_{i \in ([1:N]\backslash \{j\})  \backslash  \mathcal{A}^{\star}_j} r_{i,s}^{\prime} 
\right)
\right]
\\& \geq 
\sum_{s\in [0:1]^N} \lambda_s
\left [
\sum_{j=1}^{N-1} 
\left(\max_{i \in \mathcal{A}_{{\rm{F}}j}} \ell_{i,s}^{\prime} 
+ \max_{i \in [1:N] \backslash \mathcal{A}_{{\rm{F}}j}} r_{i,s}^{\prime} \right)
\right ]
\\& = \sum_{j=1}^{N-1} \sum_{s\in [0:1]^N} \lambda_s
\left(
\max_{i \in \mathcal{A}_{{\rm{F}}j}} \ell_{i,s}^{\prime}
+ \max_{i \in [1:N] \backslash \mathcal{A}_{{\rm{F}}j}} r_{i,s}^{\prime} 
\right)
\\& \geq (N-1)\min_{\mcal{A} \subseteq [1:N]} \left\{ \sum_{s\in [0:1]^N} \lambda_s
\left(
    \max_{i \in \mcal{A}} \ell_{i,s}^{\prime}
    + \max_{i \in [1:N] \backslash \mcal{A}} r_{i,s}^{\prime} 
\right) \right \}
\\
&= (N-1) \mathsf{R}^{\lambda}_{\mathcal{N}_{[1:N]}}.
\end{align*}
This completes the proof of Lemma~\ref{Lemma:HD3From4}.

\section{Detailed analysis for the network in~\eqref{eq:WorstNetwEx}}
\label{app:cut1}
In this section, we analyze in details the network in~\eqref{eq:WorstNetwEx}.
We start by deriving an upper bound and a lower bound on 
${\mathsf{C}}_{\mathcal{N}_{[1:N]}}$
for the network described in~\eqref{eq:WorstNetwEx} and show they are both equal to one, hence proving 
${\mathsf{C}}_{\mathcal{N}_{[1:N]}}=1$.
%
A trivial upper bound on 
${\mathsf{C}}_{\mathcal{N}_{[1:N]}}$
is given by 
${\mathsf{C}}^{{\rm{FD}}}_{\mathcal{N}_{[1:N]}}$,
i.e., 
${\mathsf{C}}_{\mathcal{N}_{[1:N]}} \leq {\mathsf{C}}^{{\rm{FD}}}_{\mathcal{N}_{[1:N]}}$.
It is not difficult to see that, for the network in~\eqref{eq:WorstNetwEx}, 
${\mathsf{C}}^{{\rm{FD}}}_{\mathcal{N}_{[1:N]}}=1$,
which implies 
${\mathsf{C}}_{\mathcal{N}_{[1:N]}}\leq1$.

We now derive a lower bound on ${\mathsf{C}}_{\mathcal{N}_{[1:N]}}$.
We start by considering even values for $N$.
Let the network in~\eqref{eq:WorstNetwEx} operate only in $2$ states with the same duration, namely,
\begin{align*}
\lambda_{\underbrace{00 \ldots 0}_{\frac{N}{2}}\underbrace{11 \ldots 1}_{\frac{N}{2}}}
=
\lambda_{\underbrace{11 \ldots 1}_{\frac{N}{2}}\underbrace{00 \ldots 0}_{\frac{N}{2}}}
=
\frac{1}{2}.
\end{align*}
In other words, half of the time the first $\frac{N}{2}$ relays listen, while the remaining $\frac{N}{2}$ relays transmit and half of the time the opposite occurs.
Let ${R}^{\rm{E}}_{\mathcal{N}_{[1:N]}}$ be the corresponding approximate achievable rate; 
clearly we have 
${\mathsf{C}}_{\mathcal{N}_{[1:N]}} \geq {R}^{\rm{E}}_{\mathcal{N}_{[1:N]}}$.
Let $\left \{ \mathcal{M}_1, \mathcal{M}_2\right \}$ be a partition of $[1:N]$, where $\mathcal{M}_1 = \left [ 1:\frac{N}{2}\right ]$. 
With this we have
\begin{align*}
{R}^{\rm{E}}_{\mathcal{N}_{[1:N]}}
&=
\min_{\mathcal{A} \subseteq [1:N]}
\left \{
\frac{1}{2} \max_{i \in \mathcal{A} \cap \mathcal{M}_1} \ell_i
+
\frac{1}{2} \max_{i \in \mathcal{A}^c \cap \mathcal{M}_2} r_i
+
\frac{1}{2} \max_{i \in \mathcal{A} \cap \mathcal{M}_2} \ell_i
+
\frac{1}{2} \max_{i \in \mathcal{A}^c \cap \mathcal{M}_1} r_i
\right \}
\\&=
\frac{1}{2}\min_{\mathcal{A} \subseteq [1:N]}
\left \{
\left [ 
\max_{i \in \mathcal{A} \cap \mathcal{M}_1} \ell_i
+
\max_{i \in \mathcal{A}^c \cap \mathcal{M}_1} r_i
\right ]
+
\left [
\max_{i \in \mathcal{A} \cap \mathcal{M}_2} \ell_i
+
\max_{i \in \mathcal{A}^c \cap \mathcal{M}_2} r_i
\right ]
\right \}
\\& \geq 
\frac{1}{2}
\left [
\min_{\mathcal{A} \subseteq [1:N]} 
\left \{
\max_{i \in \mathcal{A} \cap \mathcal{M}_1} \ell_i
+
\max_{i \in \mathcal{A}^c \cap \mathcal{M}_1} r_i
\right \}
+
\min_{\mathcal{A} \subseteq [1:N]} 
\left \{
\max_{i \in \mathcal{A} \cap \mathcal{M}_2} \ell_i
+
\max_{i \in \mathcal{A}^c \cap \mathcal{M}_2} r_i
\right \}
\right ]
\\& = \frac{1}{2} {\left( {\mathsf{C}}^{{\rm{FD}}}_{\mathcal{N}_{[1:N]}} +{\mathsf{C}}^{{\rm{FD}}}_{\mathcal{N}_{[1:N]}} \right)} =1.
\end{align*}
Hence, for even values of $N$, we have ${\mathsf{C}}_{\mathcal{N}_{[1:N]}}\geq1$, which together with the upper bound ${\mathsf{C}}_{\mathcal{N}_{[1:N]}}\leq1$, implies ${\mathsf{C}}_{\mathcal{N}_{[1:N]}} = 1$.
We now consider odd values for $N$.
Let the network in~\eqref{eq:WorstNetwEx} operate only in $2$ states with the same duration, namely
\begin{align*}
\lambda_{\underbrace{00 \ldots 0}_{\frac{N-1}{2}}\underbrace{11 \ldots 1}_{\frac{N-1}{2}} \underbrace{1}_{1}}
=
\lambda_{\underbrace{11 \ldots 1}_{\frac{N-1}{2}}\underbrace{00 \ldots 0}_{\frac{N-1}{2}} \underbrace{1}_{1}}
=
\frac{1}{2}.
\end{align*}
In other words, the $N$-th relay is always transmitting, while half of the time the first $\frac{N-1}{2}$ relays listen, while the remaining $\frac{N-1}{2}$ relays transmit and half of the time the opposite occurs.
Let ${R}^{\rm{O}}_{\mathcal{N}_{[1:N]}}$ be the corresponding approximate achievable rate; 
clearly we have ${\mathsf{C}}_{\mathcal{N}_{[1:N]}} \geq {R}^{\rm{O}}_{\mathcal{N}_{[1:N]}}$.
Let $\mathcal{M}_1=\left [ 1: \frac{N-1}{2} \right ]$ 
and 
$\mathcal{M}_2 = \left[\frac{N+1}{2}:N-1 \right ]$.
With this we have
\begin{align*}
{R}^{\rm{O}}_{\mathcal{N}_{[1:N]}}
&=
\min_{\mathcal{A} \subseteq [1:N]}
\left \{
\frac{1}{2} \max_{i \in \mathcal{A} \cap \mathcal{M}_1} \ell_i
+
\frac{1}{2} \max_{i \in \mathcal{A}^c \cap \left ( \mathcal{M}_2 \cup \{N\} \right )} r_i
+
\frac{1}{2} \max_{i \in \mathcal{A} \cap \mathcal{M}_2} \ell_i
+
\frac{1}{2} \max_{i \in \mathcal{A}^c \cap \left(\mathcal{M}_1 \cup \{N\}\right )} r_i
\right \}
\\&=
\frac{1}{2}\min_{\mathcal{A} \subseteq [1:N]}
\left \{
\left [ 
\max_{i \in \mathcal{A} \cap \mathcal{M}_1 } \ell_i
+
\max_{i \in \mathcal{A}^c \cap \left(\mathcal{M}_1 \cup \{N\}\right)} r_i
\right ]
+
\left [
\max_{i \in \mathcal{A} \cap \mathcal{M}_2} \ell_i
+
\max_{i \in \mathcal{A}^c \cap \left( \mathcal{M}_2 \cup \{N\}\right )} r_i
\right ]
\right \}
\\& \geq 
\frac{1}{2}
\left [
\min_{\mathcal{A} \subseteq [1:N]} 
\left \{
\max_{i \in \mathcal{A} \cap \mathcal{M}_1} \ell_i
+
\max_{i \in \mathcal{A}^c \cap \left(\mathcal{M}_1 \cup \{N\}\right)} r_i
\right \}
+
\min_{\mathcal{A} \subseteq [1:N]} 
\left \{
\max_{i \in \mathcal{A} \cap \mathcal{M}_2} \ell_i
+
\max_{i \in \mathcal{A}^c \cap \left(\mathcal{M}_2 \cup \{N\}\right)} r_i
\right \}
\right ]
\\& \stackrel{{\rm{(a)}}}{=} 
\frac{1}{2}
\left [
\min_{\mathcal{A} \subseteq [1:N]} 
\left \{
\max_{i \in \mathcal{A} \cap \left(\mathcal{M}_1 \cup \{N\}\right)} \ell_i
+
\max_{i \in \mathcal{A}^c \cap \left(\mathcal{M}_1 \cup \{N\}\right)} r_i
\right \}
\right.
\\& \left.
\quad +
\min_{\mathcal{A} \subseteq [1:N]} 
\left \{
\max_{i \in \mathcal{A} \cap \left(\mathcal{M}_2 \cup \{N\}\right)} \ell_i
+
\max_{i \in \mathcal{A}^c \cap \left(\mathcal{M}_2 \cup \{N\}\right)} r_i
\right \}
\right ]
 = \frac{1}{2} {\left( {\mathsf{C}}^{{\rm{FD}}}_{\mathcal{N}_{[1:N]}} +{\mathsf{C}}^{{\rm{FD}}}_{\mathcal{N}_{[1:N]}} \right)} =1,
\end{align*}
where the equality in $\rm{(a)}$ follows since the $N$-th relay is never in the minimum cut $\mathcal{A}$ as otherwise the approximate capacity would be infinity (since from~\eqref{eq:WorstNetwEx} we have $\ell_N\rightarrow \infty$).
Hence, also for odd values of $N$ we have ${\mathsf{C}}_{\mathcal{N}_{[1:N]}}\geq1$, which together with the upper bound ${\mathsf{C}}_{\mathcal{N}_{[1:N]}}\leq1$, implies ${\mathsf{C}}_{\mathcal{N}_{[1:N]}}=1$.
This concludes the proof that ${\mathsf{C}}_{\mathcal{N}_{[1:N]}}=1$ for the network in~\eqref{eq:WorstNetwEx}.

Now, assume that $N = 4t-2$, where $t \in \mathbb{N} \backslash \{0\}$ and with this
suppose we want to select the best subnetwork $\mathcal{N}_{\mathcal{K}}$ with $|\mathcal{K}|=1$ in the network $\mathcal{N}_{[1:N]}$ in~\eqref{eq:WorstNetwEx}, i.e., we want to select the best relay.
From~\eqref{eq:singleCapGen} we obtain that the approximate single capacity of the $i$-th relay with $i \in \left [1:\left \lfloor \frac{N}{2} \right \rfloor \right]$ is given by
\begin{subequations}
\label{eq:singleCap}
\begin{align}
    & \mathsf{C}_{\mcal{N}_{\{i\}}} = \mathsf{C}_{\mcal{N}_{\left \{\left \lfloor N/2 \right \rfloor +i\right \}}}= \frac{\ell_i r_i}{\ell_i + r_i} = \frac{2i \left( N-2i+2 \right)}{N \left( N+2\right)},
\label{eq:singleCap1}
\\& \text{if }N\text{ is odd:} \ \ \mathsf{C}_{\mcal{N}_{\{N\}}} = \frac{1}{N}.
\label{eq:singleCap2}
\end{align}
\end{subequations}
It is not difficult to see 
that the expression of $\mathsf{C}_{\mathcal{N}_{\{i\}}}$ in~\eqref{eq:singleCap} achieves its maximum value for
\begin{align}
\label{eq:iStar}
i^{\star} = \frac{N+2}{4},
\end{align}
for which
\begin{align}
\label{eq:istar}
 \mathsf{C}_{\mcal{N}_{\{i^{\star}\}}} &=  \frac{2  \frac{N+2 }{4}  \left( N-2   \frac{N+2 }{4} +2 \right)}{N \left( N+2\right)} = \frac{\frac{N+2}{2} \left( \frac{2N+4}{4} \right)}{N (N+2)} 
= \frac{N+2}{4N}
= \frac{t}{4t-2},
\end{align}
which for $t \rightarrow \infty$ gives
\begin{align*}
\mathsf{C}_{\mcal{N}_{\{i^{\star}\}}}{=} \frac{1}{4} \implies 
\mathsf{C}_{\mathcal{N}_{\mathcal{K}}} {=} \frac{1}{4} {\mathsf{C}}_{\mathcal{N}_{[1:N]}}, \ |\mathcal{K}|=1.
\end{align*}
Now, for the same network, suppose we want to select the best subnetwork $\mathcal{N}_{\mathcal{K}}$ with $|\mathcal{K}|=2$, i.e., we want to select the best $2$-relay subnetwork.
Clearly from Lemma~\ref{lem:part} (partition lemma), if we select relays number $i \in [1:N]$ and $j \in [1:N]$ with $i\neq j$ a trivial upper bound on the approximate capacity $\mathsf{C}_{\mcal{N}_{\{i,j\}}}$ is given by
\begin{align*}
    \mathsf{C}_{\mcal{N}_{\{i,j\}}} \leq \mathsf{C}_{\mcal{N}_{\{i\}}} + \mathsf{C}_{\mcal{N}_{\{j\}}}
\leq
2 \mathsf{C}_{\mcal{N}_{\{i^{\star}\}}}.
\end{align*}
Consider relays number $i^{\star}$ and $j^{\star}=i^{\star}+\frac{N}{2}$, where $i^\star$ is defined in~\eqref{eq:iStar}.
By substituting $j^{\star}$ into~\eqref{eq:WorstNetwEx} we obtain
\begin{align*}
& \ell_{i^{\star}} = \ell_{j^{\star}} = r_{i^{\star}} = r_{j^{\star}} = \frac{2  \frac{N+2}{4} }{N},
\end{align*}
which implies $\mathsf{C}_{\mcal{N}_{\{i^{\star}\}}} = \mathsf{C}_{\mcal{N}_{\{j^{\star}\}}}$, where $\mathsf{C}_{\mcal{N}_{\{i^{\star}\}}}$ is defined in~\eqref{eq:istar} and from~\cite{BagheriIT2014} we have
\begin{align}
\label{eq:Cij}
\mathsf{C}_{\mcal{N}_{\{i^{\star}, j^{\star}\}} }
= \mathsf{C}_{\mcal{N}_{\{i^{\star}\}}}+\mathsf{C}_{\mcal{N}_{\{j^{\star}\}} }
= 2 \mathsf{C}_{\mcal{N}_{\{i^{\star}\}}}
= \frac{t}{2t-1},
\end{align}
which for $t \rightarrow \infty $ gives
\begin{align*}
\mathsf{C}_{\mcal{N}_{\{i^{\star}, j^{\star}\}} } {=} \frac{1}{2} \implies 
\mathsf{C}_{\mathcal{N}_{\mathcal{K}}} {=} \frac{1}{2} {\mathsf{C}}_{\mathcal{N}_{[1:N]}}, \ |\mathcal{K}|=2.
\end{align*}
So, the network in~\eqref{eq:WorstNetwEx}, for $N = 4t-2$, where $t \in \mathbb{N} \backslash \{0\}$,
represents an example for the network described in the statement of Theorem~\ref{thm:highN}. 
This concludes the proof of Theorem~\ref{thm:highN}.

\end{appendices}

\bibliographystyle{IEEEtran}
\bibliography{ITjournalBib.bib}

\begin{thebibliography}{10}
\providecommand{\url}[1]{#1}
\csname url@samestyle\endcsname
\providecommand{\newblock}{\relax}
\providecommand{\bibinfo}[2]{#2}
\providecommand{\BIBentrySTDinterwordspacing}{\spaceskip=0pt\relax}
\providecommand{\BIBentryALTinterwordstretchfactor}{4}
\providecommand{\BIBentryALTinterwordspacing}{\spaceskip=\fontdimen2\font plus
\BIBentryALTinterwordstretchfactor\fontdimen3\font minus
  \fontdimen4\font\relax}
\providecommand{\BIBforeignlanguage}[2]{{%
\expandafter\ifx\csname l@#1\endcsname\relax
\typeout{** WARNING: IEEEtran.bst: No hyphenation pattern has been}%
\typeout{** loaded for the language `#1'. Using the pattern for}%
\typeout{** the default language instead.}%
\else
\language=\csname l@#1\endcsname
\fi
#2}}
\providecommand{\BIBdecl}{\relax}
\BIBdecl

\bibitem{NazarogluIT2014}
C.~Nazaroglu, A.~{\"O}zg{\"u}r, and C.~Fragouli, ``Wireless network
  simplification: The {G}aussian {N}-relay diamond network,'' \emph{IEEE
  Transactions on Information Theory}, vol.~60, no.~10, pp. 6329--6341, October
  2014.

\bibitem{coveElGamal}
T.~Cover and A.~El~Gamal, ``Capacity theorems for the relay channel,''
  \emph{IEEE Transactions on Information Theory}, vol.~25, no.~5, pp. 572 --
  584, September 1979.

\bibitem{LimIT2011}
S.~Lim, Y.-H. Kim, A.~El~Gamal, and S.-Y. Chung, ``Noisy network coding,''
  \emph{IEEE Transactions on Information Theory}, vol.~57, no.~5, pp. 3132
  --3152, May 2011.

\bibitem{AvestimehrIT2011}
A.~S. Avestimehr, S.~N. Diggavi, and D.~N.~C. Tse, ``Wireless network
  information flow: A deterministic approach,'' \emph{IEEE Transactions on
  Information Theory}, vol.~57, no.~4, pp. 1872--1905, April 2011.

\bibitem{OzgurIT2013}
A.~{\"O}zg{\"u}r and S.~N. Diggavi, ``Approximately achieving {G}aussian relay
  network capacity with lattice-based qmf codes,'' \emph{IEEE Transactions on
  Information Theory}, vol.~59, no.~12, pp. 8275--8294, December 2013.

\bibitem{LimISIT2014}
S.~H. Lim, K.~T. Kim, and Y.~H. Kim, ``Distributed decode-forward for
  multicast,'' in \emph{IEEE International Symposium on Information Theory
  (ISIT)}, June 2014, pp. 636--640.

\bibitem{CardoneIT2014}
M.~Cardone, D.~Tuninetti, R.~Knopp, and U.~Salim, ``Gaussian half-duplex relay
  networks: improved constant gap and connections with the assignment
  problem,'' \emph{IEEE Transactions on Information Theory}, vol.~60, no.~6,
  pp. 3559 -- 3575, June 2014.

\bibitem{KramerAllerton2004}
G.~Kramer, ``Models and theory for relay channels with receive constraints,''
  in \emph{42nd Annual Allerton Conference on Communication, Control, and
  Computing}, September 2004, pp. 1312--1321.

\bibitem{SenguptaITW2012}
A.~Sengupta, I.-H. Wang, and C.~Fragouli, ``Optimizing quantize-map-and-forward
  relaying for {G}aussian diamond networks,'' in \emph{IEEE Information Theory
  Workshop (ITW)}, September 2012, pp. 381--385.

\bibitem{ChernITW2012}
B.~Chern and A.~{\"O}zg{\"u}r, ``Achieving the capacity of the n-relay
  {G}aussian diamond network within logn bits,'' in \emph{IEEE Information
  Theory Workshop (ITW)}, September 2012, pp. 377--380.

\bibitem{OzgurISIT2015}
T.~A. Courtade and A.~{\"O}zg{\"u}r, ``Approximate capacity of {G}aussian relay
  networks: Is a sublinear gap to the cutset bound plausible?'' in \emph{IEEE
  International Symposium on Information Theory (ISIT)}, June 2015, pp.
  2251--2255.

\bibitem{OzgurAllerton2015}
X.~Wu and A.~{\"O}zg{\"u}r, ``Cut-set bound is loose for {G}aussian relay
  networks,'' in \emph{53rd Annual Allerton Conference on Communication,
  Control, and Computing}, October 2015, pp. 1135--1142.

\bibitem{CardoneITW2015}
M.~Cardone, D.~Tuninetti, and R.~Knopp, ``The approximate optimality of simple
  schedules for half-duplex multi-relay networks,'' in \emph{IEEE Information
  Theory Workshop (ITW)}, April 2015, pp. 1--5.

\bibitem{BrahmaISIT2012}
S.~Brahma, A.~{\"O}zg{\"u}r, and C.~Fragouli, ``Simple schedules for
  half-duplex networks,'' in \emph{IEEE International Symposium on Information
  Theory (ISIT)}, July 2012, pp. 1112--1116.

\bibitem{BagheriIT2014}
H.~Bagheri, A.~Motahari, and A.~Khandani, ``On the capacity of the half-duplex
  diamond channel under fixed scheduling,'' \emph{IEEE Transactions on
  Information Theory}, vol.~60, no.~6, pp. 3544--3558, June 2014.

\bibitem{BrahmaISIT2014}
S.~Brahma and C.~Fragouli, ``Structure of optimal schedules in diamond
  networks,'' in \emph{IEEE International Symposium on Information Theory
  (ISIT)}, June 2014, pp. 641--645.

\bibitem{BrahmaIT2016}
S.~Brahma, C.~Fragouli, and A.~{\"O}zg{\"u}r, ``On the complexity of scheduling
  in half-duplex diamond networks,'' \emph{IEEE Transactions on Information
  Theory}, vol.~62, no.~5, pp. 2557--2572, May 2016.

\bibitem{EzzeldinISIT2017}
Y.~H. Ezzeldin, M.~Cardone, C.~Fragouli, and D.~Tuninetti, ``Finding simple
  half-duplex schedules in {G}aussian relay line networks,'' \emph{to appear in
  IEEE International Symposium on Information Theory (ISIT)}, June 2017.

\bibitem{SenguptaITW2014}
S.~Brahma, A.~Sengupta, and C.~Fragouli, ``Switched local schedules for diamond
  networks,'' in \emph{IEEE Information Theory Workshop (ITW)}, November 2014,
  pp. 656--660.

\bibitem{EtkinParvareshShomoronyAvestimehr}
R.~Etkin, F.~Parvaresh, I.~Shomorony, and A.~Avestimehr, ``Computing
  half-duplex schedules in {G}aussian relay networks via min-cut
  approximations,'' \emph{IEEE Transactions on Information Theory}, vol.~60,
  no.~11, pp. 7204--7220, November 2014.

\bibitem{EzzeldinISIT2016}
Y.~H. Ezzeldin, A.~Sengupta, and C.~Fragouli, ``Wireless network
  simplification: Beyond diamond networks,'' in \emph{IEEE International
  Symposium on Information Theory (ISIT)}, July 2016, pp. 2594--2598.

\bibitem{BrahmaISIT2014Relay}
S.~Brahma and C.~Fragouli, ``A simple relaying strategy for diamond networks,''
  in \emph{IEEE International Symposium on Information Theory (ISIT)}, June
  2014, pp. 1922--1926.

\bibitem{HoAllerton2010}
T.~Ho, M.~Effros, and S.~Jalali, ``On equivalence between network topologies,''
  in \emph{48th Annual Allerton Conference on Communication, Control, and
  Computing}, September 2010, pp. 391--398.

\bibitem{JalaliITA2011}
S.~Jalali, M.~Effros, and T.~Ho, ``On the impact of a single edge on the
  network coding capacity,'' in \emph{Information Theory and Applications
  Workshop (ITA)}, February 2011, pp. 1--5.

\bibitem{EtkinIT2014}
F.~Parvaresh and R.~Etkin, ``Efficient capacity computation and power
  optimization for relay networks,'' \emph{IEEE Transactions on Information
  Theory}, vol.~60, no.~3, pp. 1782--1792, March 2014.

\end{thebibliography}

\end{document}